\documentstyle[epsfig]{aipproc}
\begin{document}
\title{Recent results on Charm Physics from Fermilab\thanks{This 
work is partially supported by CLAF/CNPq Brazil and CONACyT M\'exico}
\thanks{To appear, Proc. {\it VII Mexican Workshop on Particles 
and Fields}, Merida Yuc. M\'ex., Nov.1999} }

\author{J. C. Anjos\thanks{janjos@lafex.cbpf.br} and 
E. Cuautle\thanks{ecuautle@lafex.cbpf.br}}

\address{Centro Brasileiro de Pesquisas F\'{\i}sicas, CBPF\\
Rua Dr. Xavier Sigaud 150, 22290-180 Rio de Janeiro Brazil}. 

\maketitle

\begin{abstract}
New high statistics, high resolution fixed target experiments producing
$10^5$ - $10^6$ fully reconstructed charm particles are allowing a
detailed study of the charm sector. Recent results on charm quark
production from Fermilab fixed target experiments
E-791, SELEX and FOCUS are presented. 
\end{abstract}

\section*{Introduction}

This review contains recent results from three Fermilab fixed target
experiments dedicated to study charm physics: E791 and E871 (SELEX) of
charm hadroproduction and E831 (FOCUS) of charm photoproduction.\\
The experiment E791 used a 500 GeV/c  $\pi^-$   beam incident on
platinum and diamond target foils and took data in 1991-1992. A loose
transverse energy trigger was used to record $2\time10^{10}$ interactions.
Silicon microstrip detectors (6 in the beam and 17 downstream of the target)
provided precision track and vertex reconstruction.
The precise location of production and decay vertices of long lived charm 
particles allowed to reconstruct over $2\time10^5$ charm particles.\\
SELEX used a 600 GeV hyperon beam of negative polarity to make a mixed
beam of $\Sigma^-$ and $\pi^-$ in roughly equal numbers. The positive beam
was composed of 92 $\%$ of protons, and 8 $\%$ of $\pi^+$.
Interactions occurred in a segmented target of 5 foils, 2 Cu and 3 C.
The  experiment was designed to study charm
production in the forward hemisphere, with good mass and decay vertex
resolution for charm momenta in the range 100-500 GeV/c. A major
innovation was the use of online selection criteria
 to identify events that had evidence for a secondary vertex.  Data taking
finished in 1997.\\
The FOCUS experiment used a photon beam with $<E_{\gamma}>=170$ GeV on a
Beryllium Oxide segmented target. Sucessor of Fermilab E687, FOCUS had
upgrades in vertexing, Cerenkov identification, electromagnetic calorimetry
and muon identification. The target
 segmentation and the use of additional silicon microstrip detectors after
each pair of target segments were major improvements. The experiment took
data in 1996-1997 and collected over 7 billion triggers of charm
candidates. FOCUS has the largest charm sample available in the world,
with over $10^6$ fully reconstructed charmed particles.\\

The data available from these three experiments will allow to make high
precision charm physics, to study rare decays and to search for new
physics.

\section*{Charm production and cross sections for D mesons}

\begin{figure}[!ht] 
\begin{center}
\begin{tabular}{cccc}
\epsfig{file=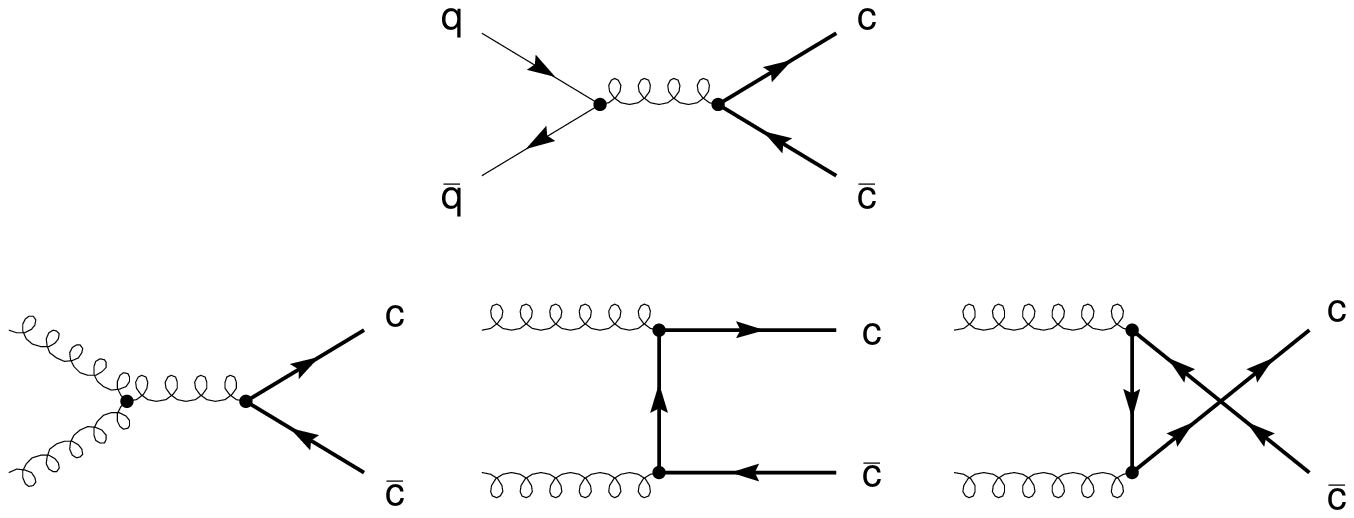,height=2.in,width=2.4in} & & &
\epsfig{file=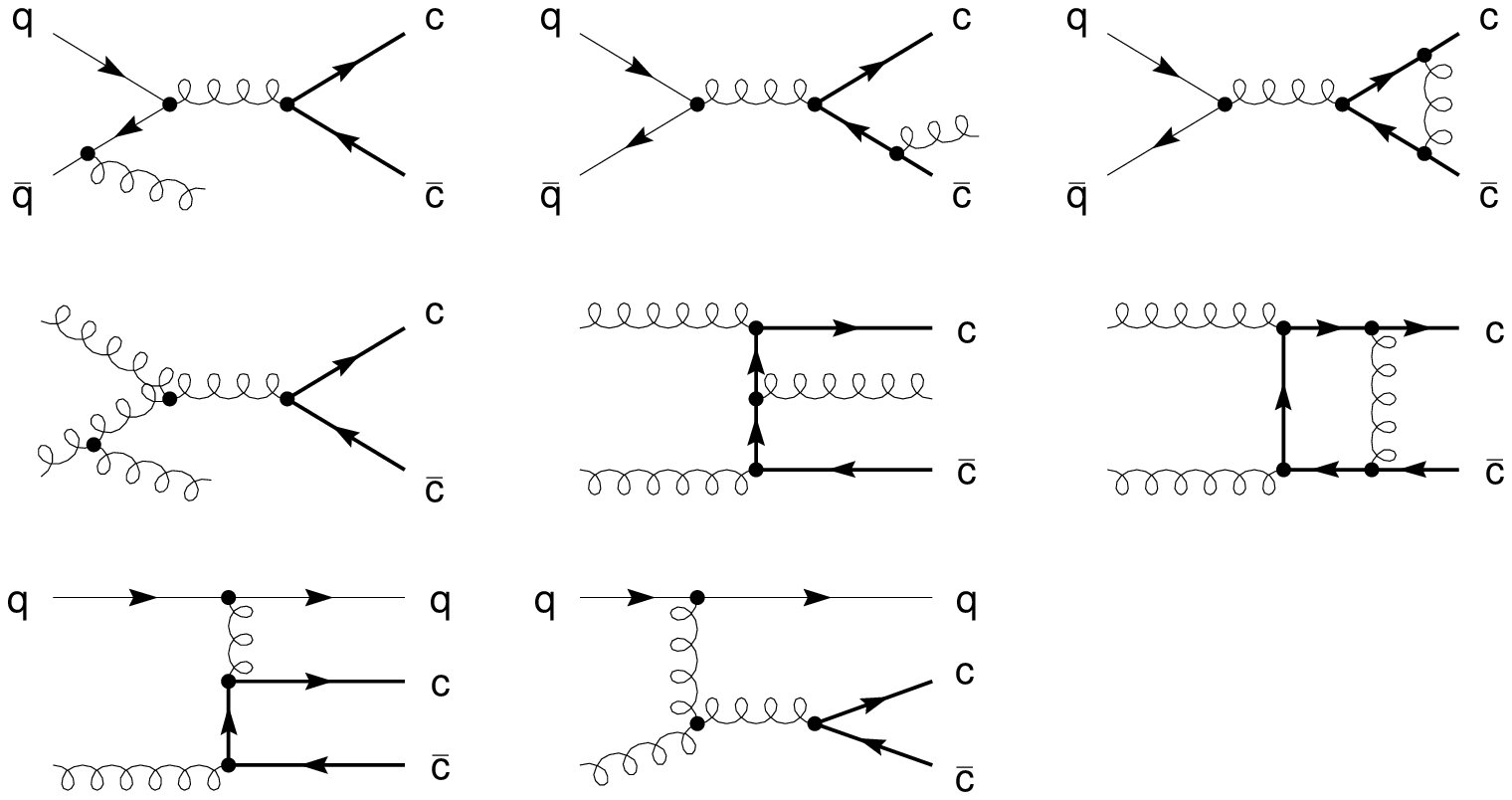,height=2.4in,width=2.8in}\\
\end{tabular}
\end{center}
\vspace{10pt}
\caption{Some  of Feynman diagrams for charm hadroproduction: leading order 
(left) and  Next-to-Leading Order (right side)}
\label{fig1}
\end{figure}

Charm production is a combination of short and long range processes.
Perturbative Quantum Chromodynamics (pQCD) can be used to calculate the
parton cross section, the short range process. The long range process of
parton hadronization has to be modeled from experimental data. The two
processes occur at different time scales and should not affect each
other, leading to factorization properties. The general expression for
production of charm in hadroproductions is \cite{crossection}

\begin{equation}
\sigma(P_A,P_B)= \sum_{i,j}  \int dx_A dx_B f_i^A(x_A, \mu) 
f_j^B(x_B, \mu) \hat{\sigma}_{i,j}(\alpha_s(\mu), x_A P_A, x_B P_B)
\label{eq1}
\end{equation}
\noindent
where $f_i^A(x_A,\mu)$ is the probability of finding parton type $i$
inside the hadron $A$ with momentum fraction $x_A$, and $\mu$ is the scale at
which the process occurs. The  other hadron taking part in the interaction
has $f_j^B(x_B, \mu)$. The elementary parton cross section
$\hat{\sigma}_{i,j}(\alpha_s(\mu), x_A P_A, x_B P_B)$ term is calculated
by QCD, according to parton models. Although QCD is well defined theory,
solutions to most problems are quite difficult. So, to calculate the parton
cross section we usually do a perturbative series expansion in terms of
the strong interaction coupling constant, $\alpha_s$, and calculate at
leading order (LO), next-to-leading order (NLO) and so on. Nowadays most 
calculations include NLO. Some Feynman diagrams 
for hadroproduction of charm at LO and NLO are shown in  Fig.\ref{fig1} 
(more details can be found in  reference \cite{crossection}).

\begin{figure}[!ht] 
\begin{center}
\begin{tabular}{ccccc}
\epsfig{file=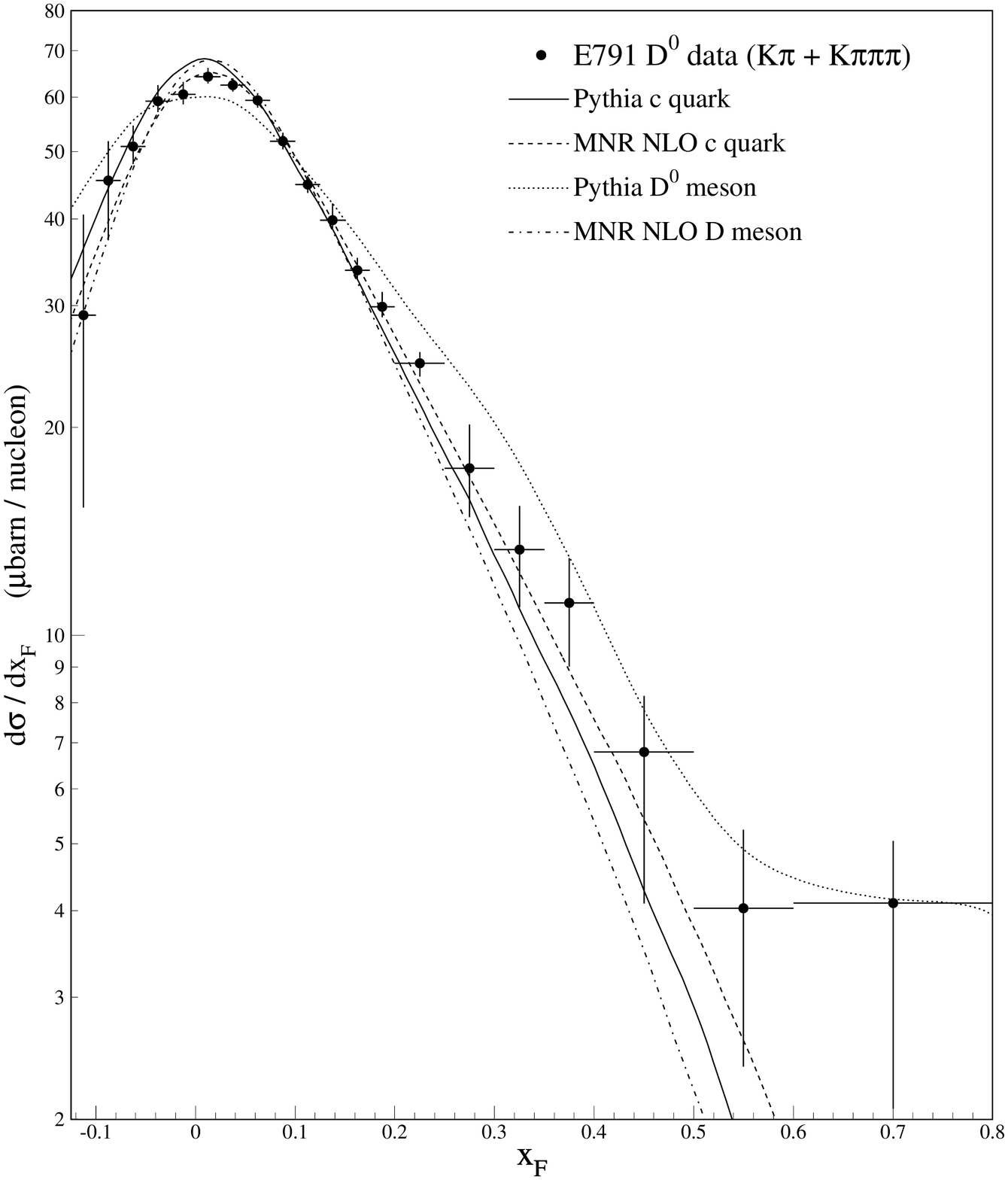,height=2.51in,width=1.5in}& & 
\epsfig{file=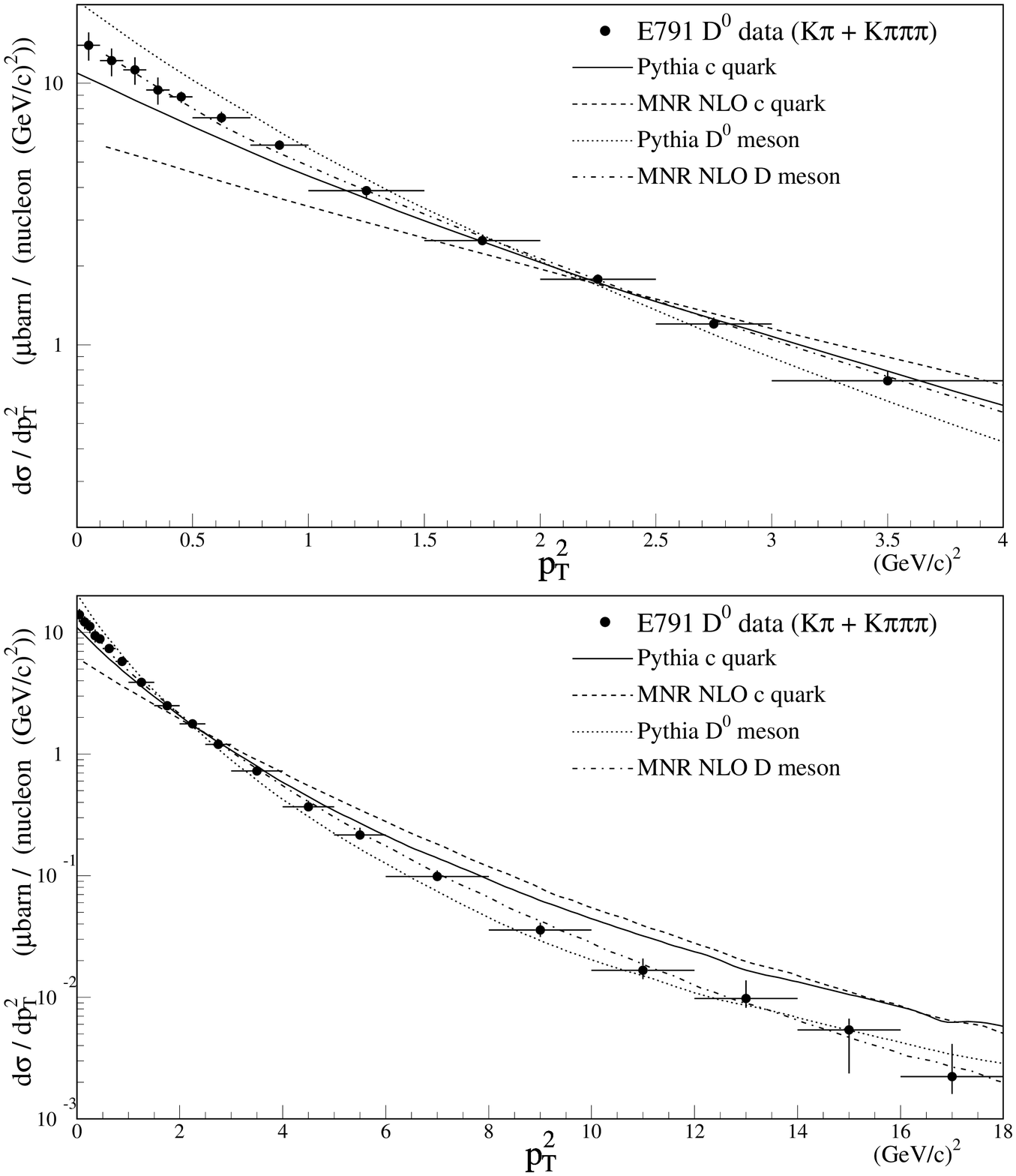,height=2.51in,width=2.in}& &
\epsfig{file=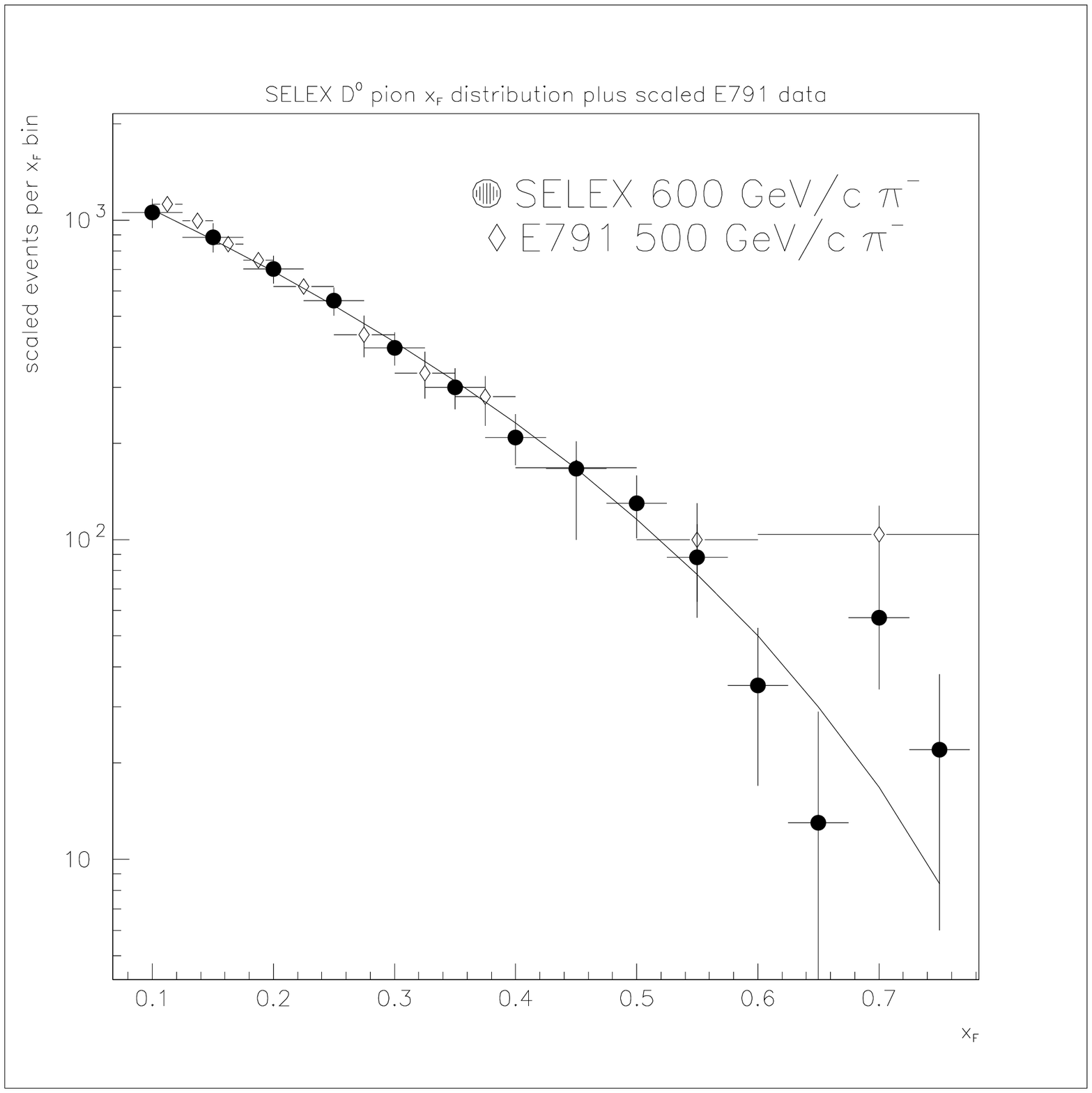,height=2.41in,width=1.3in}\\
\end{tabular}
\end{center}
\vspace{10pt}
\caption{$D^0$ production: from E791 compared 
with theoretical models as a function of $x_F$ (left), $p_t^2$ (central
column) and compared with preliminary results from SELEX (right).} 
\label{fig2}
\end{figure}

\noindent
Usually  the differential cross sections is measured as a function of the
scaled longitudinal momentum (Feynman $x_F$) $d \sigma/d x_F$, and as a
function of the transverse momentum squared $d \sigma/d p_t^2$. A
phenomenological parametrization of the double differential cross section
often used to describe the experimental data is given by,

\begin{equation}
\frac{d\sigma}{d x_F d p_t^2} = A \: (1-x_F)^n \:exp(-b p_t^2)
\label{eq2}
\end{equation}

\noindent
where $A\:, \:n$ and $b$ are  parameters used to fit the data. In fact 
kinematic considerations provide predictions for the power $n$ at high $x_F$
and QCD calculations give average $p_t$ values comparable to the charm quark 
mass.\\
E791 has measured the total forward cross section and the differential
cross sections as functions of $x_F$ and $p_t^2$ from a sample
of $88990 \pm 460$ $D^0$ mesons. Fig. \ref{fig2} shows the differential
distributions and its comparison to theoretical predictions from QCD
calculations at NLO \cite{mangano} 
and to the Monte Carlo event generator,
Pythia/Jetset \cite{pythia} for $c \bar c$ and $D \bar D$ production. Hadron 
distributions are softer than $c\bar c$
distributions due to fragmentation. With suitable choice for the intrinsic 
transverse momentum $k_t$ of the incoming partons ($k_t =1$ GeV/c), the 
Peterson fragmentation function  parameter ($\varepsilon =0.01$), and the 
charm quark 
mass $m_c =1.5$ GeV/c$^2$,  NLO $D$ meson calculation provides a
good match to the $p_t^2$ distribution and fair match to the $x_F$
distribution. The hadronization scheme implemented in Pythia/Jetset can be
adjusted to fit the data.  The large number of $D^0$ events make it
possible to clearly observe a turnover point greater than zero ($x_c =
0.0131 \pm 0.0038$) in the $x_F$ distribution. The positive value provide 
evidence  that the gluon
distribution in the pion is harder than the gluon distribution in the
nucleon \cite{cross-D0}.  The total forward neutral D meson cross section
measured by E791 is 
$\sigma (x_F > 0) = 15.4 \pm ^{1.8}_{2.3}\:\mu b$/nucleon.\\
SELEX has also preliminary data on $D^0$ cross section. Fig. \ref{fig2} 
(right) 
shows the $x_F$ distribution for $D^0$ + c.c. from E791 \cite{cross-D0} and
the preliminary data from SELEX. Up to $x_F = 0.5$ data overlap and agree
well. However SELEX has no strong evidence for rise at large $x_F$ as seen
in the E791 data. The continuous line is a fit to the data points with the
phenomenological parametrization in Eq. \ref{eq2}.

\subsection*{Correlations in the production of Charm pairs}

Observation of both of the charm particles in a hadron-produced event  can 
give additional information on the production process and allow to test
QCD predictions since both longitudinal and
transverse momenta of the charm particles and angular correlations are 
explicitly measured.\\

In the simplest parton model the charm and anticharm particles are
expected to be produced in opposite directions in the transverse plane.
However if one assumes that the incoming partons have an intrinsic transverse 
momentum, $k_t$, this will affect the transverse momentum of the heavy quark 
pair, its azymuthal correlations and the transverse distribution of a single 
quark. The partons entering the hard interaction are
indeed supposed to have a non vanishing primordial $k_t$, seen as a
nonperturbative Fermi motion of partons inside the incoming hadrons.
Typical values of $k_t$ should thus be 300-400 MeV. However it has been
noted \cite{pairs-correlation} that much higher values of $k_t$ are required, 
at or above 1 GeV, to reproduce charm data. This could be an indications of 
the importance of next  to leading and higher order effects, by which emitted 
gluons would further modify the nearly back-to-back \cite{Jeff-Sjostrand} 
production of the final charm hadrons.

E791 has measured correlations between D and $\bar D$ mesons from $791 \pm
44$ fully reconstructed charm meson pairs \cite{correlation-E791}. The main 
variables used to describe charm pair correlations are the beam direction, 
$p_t$, and either the rapidity $y$ or the Feynman scaling variable $x_F$,
and also the azymuthal distribution $\phi$. The same way the difference
and sum between these variables for two charm mesons ($D, \bar D$) $\Delta
x_F$, $\Sigma x_F$ and so on.\\
The measured  distributions are compared to predictions of the fully 
differential NLO calculation for  $c \bar c$ production  
\cite{mangano}, as well as to predictions from the Pythia/Jetset 
Monte Carlo event generator for  $c \bar c$ and $D \bar D$ production. Default 
parameter have been used for the theoretical models.\\

\begin{figure}[!ht] 
\centerline{\epsfig{file=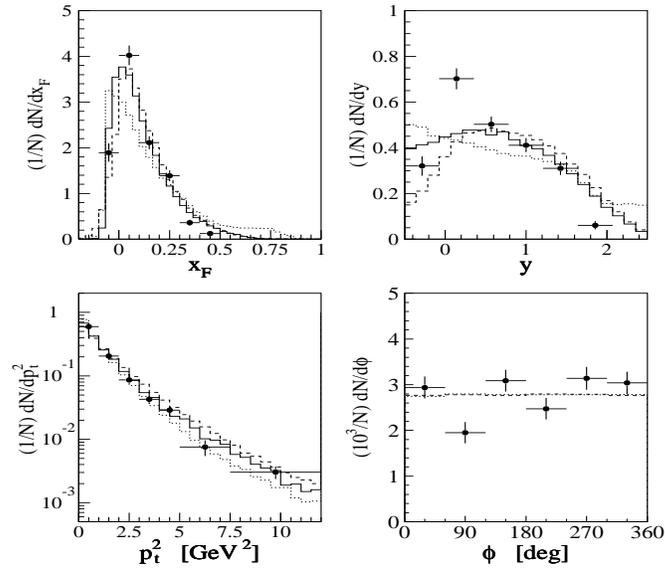,height=3.in,width=3.5in}}
\vspace{10pt}
\caption{Single-charm distributions for
$x_F$, $y$, $p_t^2$ and $\phi$:   weighted data ($\bullet$) 
NLO QCD prediction (------); {\sc Pythia/Jetset} charm quark prediction 
(${\scriptscriptstyle -\;\!-\;\!-\;\!-}$);  and {\sc Pythia/Jetset} $D$ 
meson prediction (${\scriptstyle \cdots\cdots\cdots}$). All distributions 
are obtained by summing the charm and anticharm  distributions from 
charm-pair events.} 
\label{fig9}
\end{figure}

For the single charm distributions shown in Fig.\ref{fig9}, we observe that 
for the longitudinal momentum distributions $x_F$ and $y$ the experimental 
results and theoretical predictions do not agree. In this comparison the 
experimental distributions are most similar to the NLO and Pythia/Jetset 
$c \bar c$ distributions, but are narrower than all three theoretical 
predictions. The experimental $p_t^2$ distribution
agrees quite well with all three theoretical distributions.  As expected,
both the theoretical and experimental $\phi$ distributions are consistent
with being flat.

\begin{figure}[!ht] 
\centerline{\epsfig{file=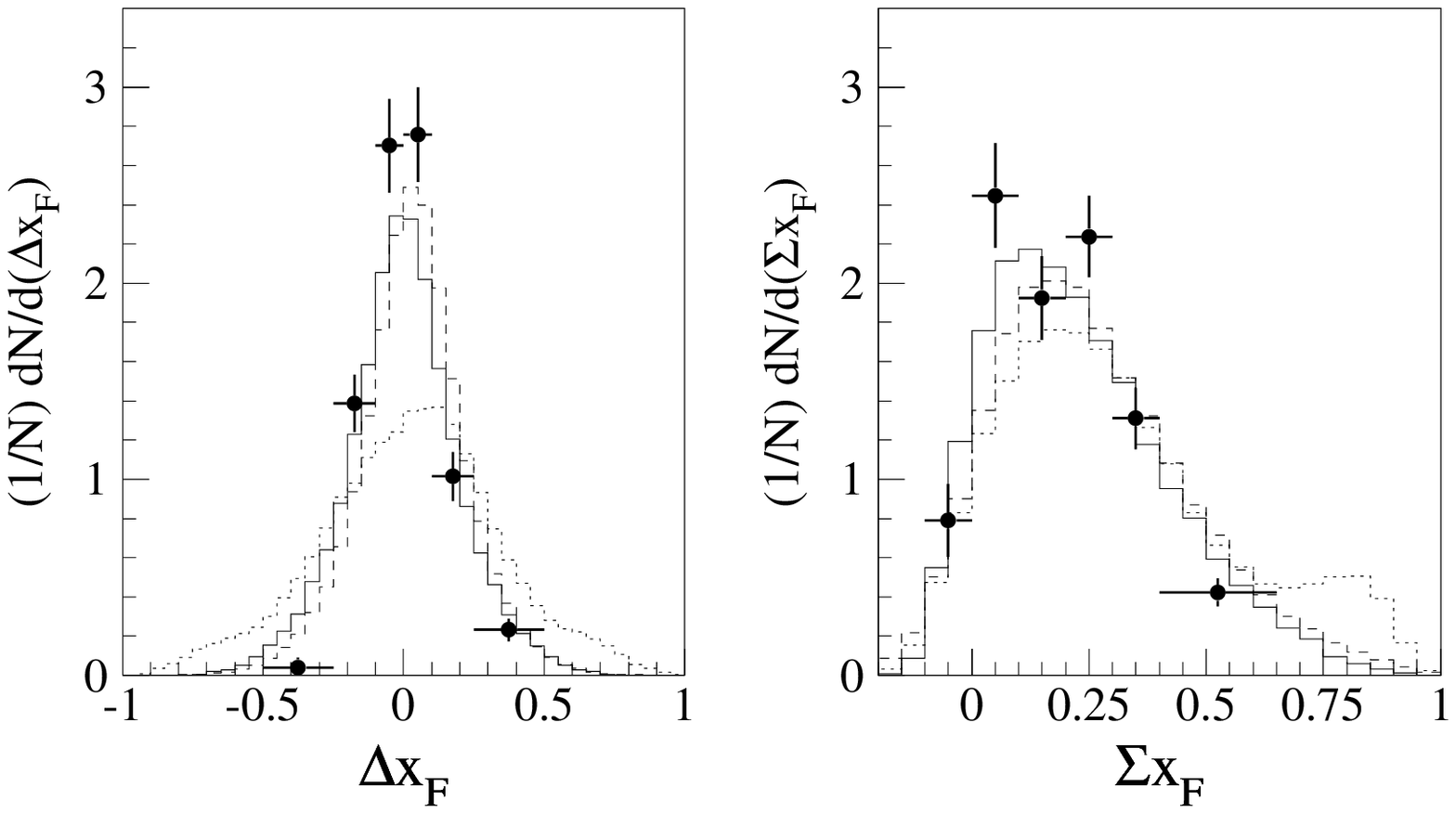,height=1.5in,width=3.5in}}
\centerline{\epsfig{file=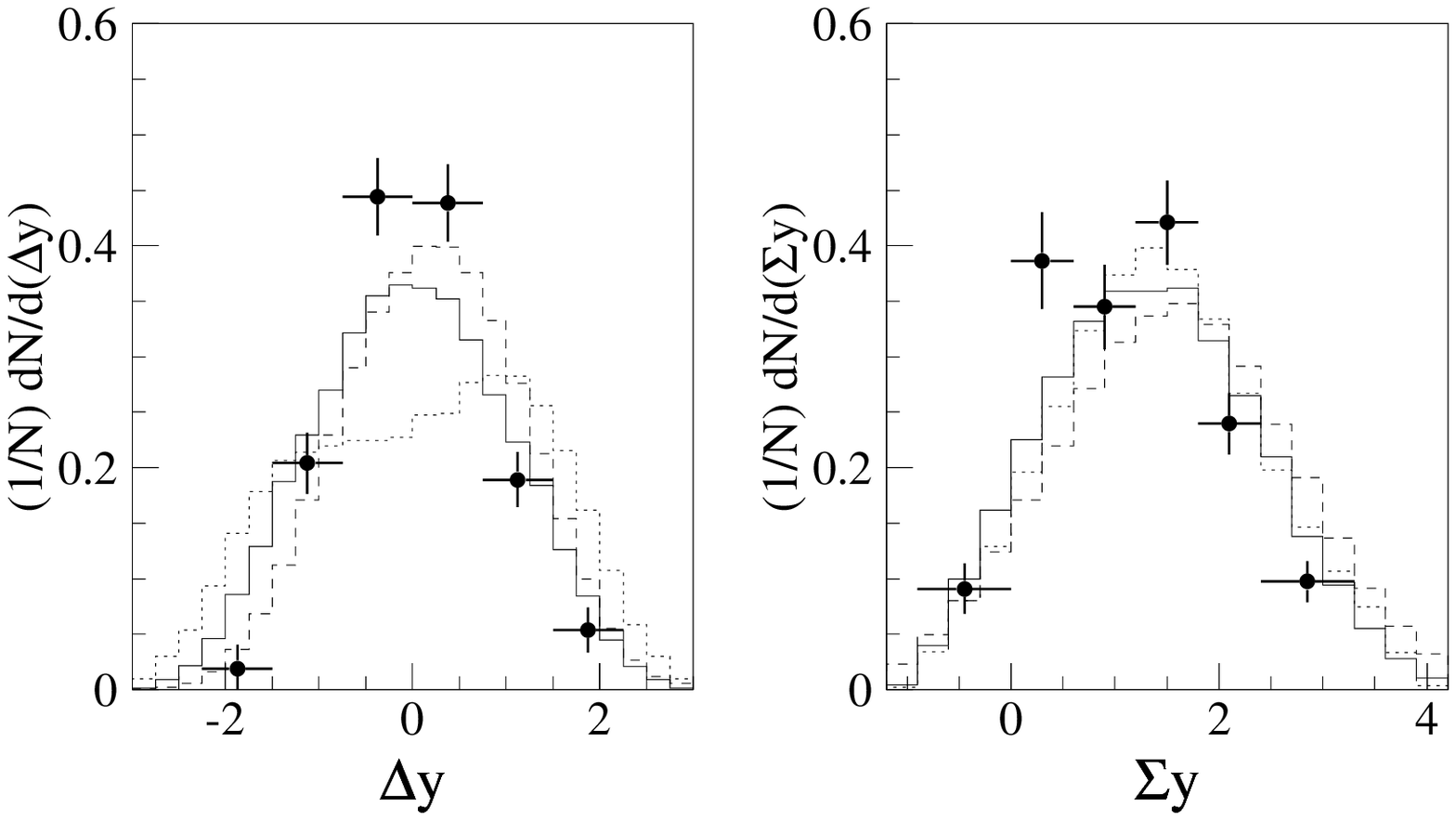,height=1.5in,width=3.5in}} 
\centerline{\epsfig{file=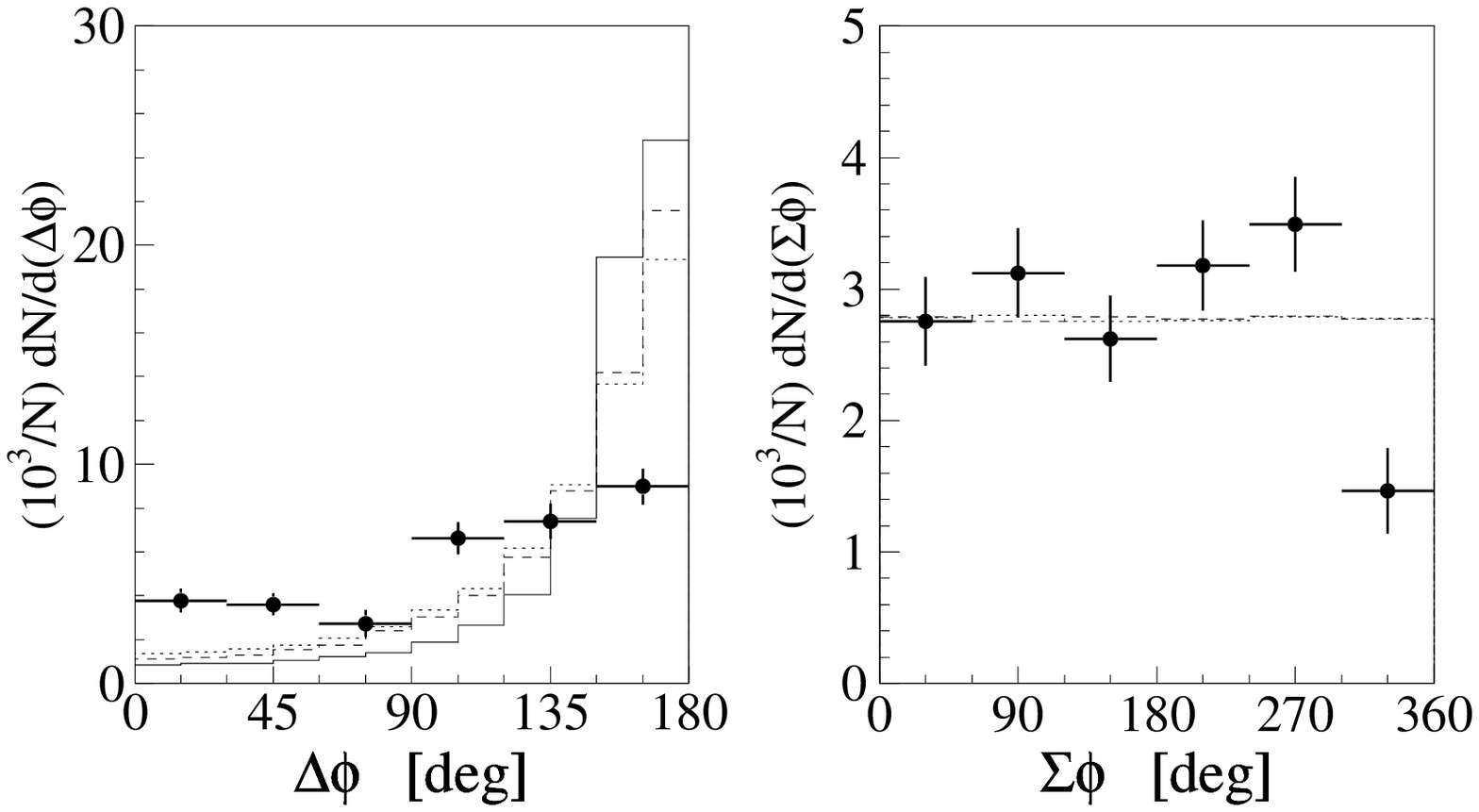,height=1.5in,width=3.5in}}
\centerline{\epsfig{file=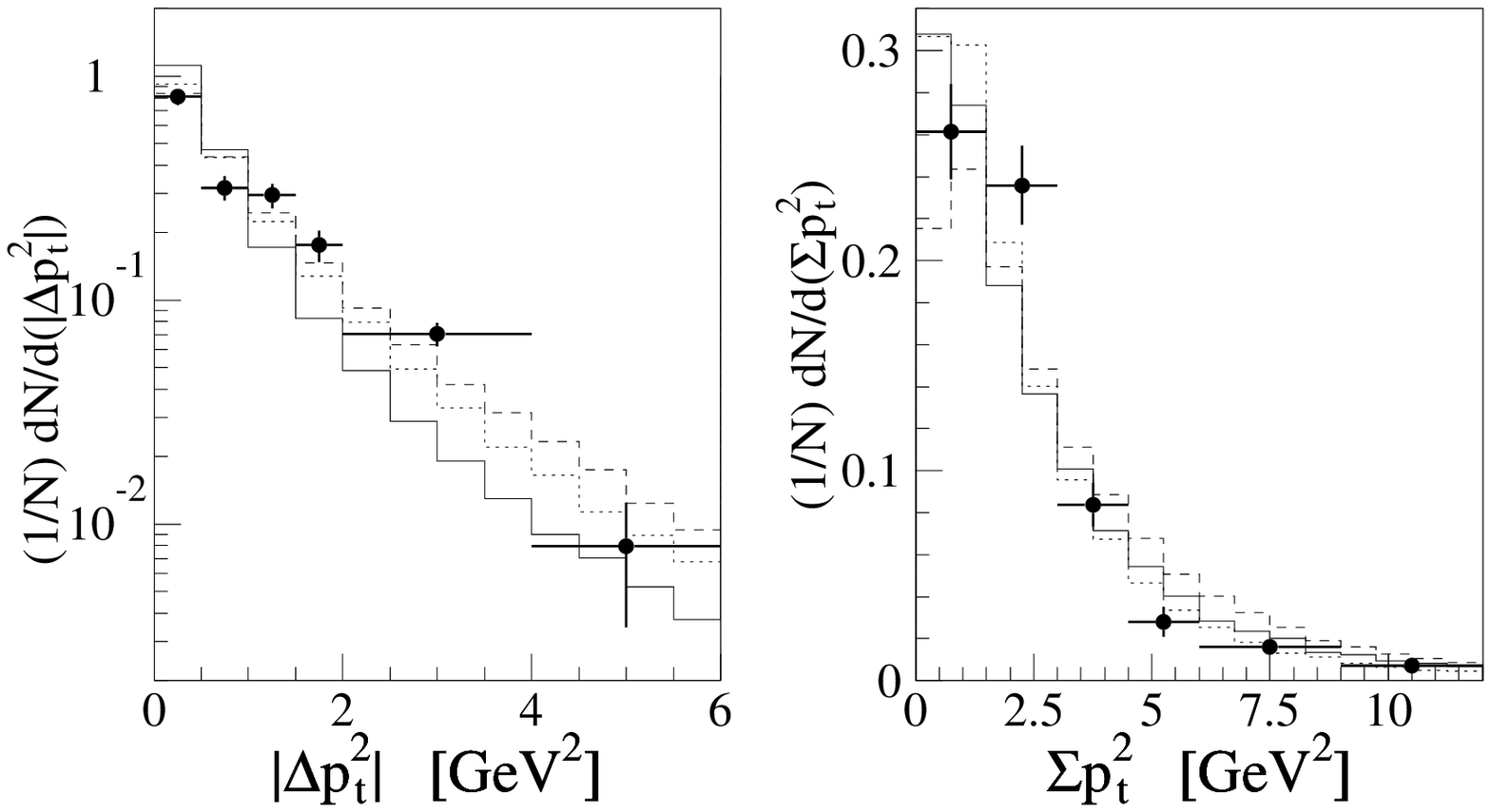,height=1.5in,width=3.5in}}
\vspace{10pt}
\caption{Charm-pair $\Delta x_F$, $\Sigma x_F$, $\Delta y$, $\Sigma y$, 
$\Delta \phi$, $\Sigma \phi$, $\Delta p_t^2$ and $\Sigma p_t^2$ distributions; 
        weighted data ($\bullet$);
        NLO QCD prediction (------);
        {\sc Pythia/Jetset}
        charm quark prediction (${\scriptscriptstyle -\;\!-\;\!-\;\!-}$);
        and {\sc Pythia/Jetset} $D$ meson prediction
        (${\scriptstyle \cdots\cdots\cdots}$).}
\label{fig10}
\end{figure}

The experimental and theoretical longitudinal distributions for pairs  
 $\Delta x_F$ , $\Sigma x_F$,
$\Delta y$ and $\Sigma y$  are shown in Fig. \ref{fig10}. As
with the single-charm distributions, the experimental results are much
closer to the two $c \bar c$ predictions than to the Pythia/Jetset $D \bar D$
predictions, but narrower than all three theoretical predictions. For the 
transverse distributions for charm pairs any observed discrepancy between data 
and theory must derive from the theory modeling the correlations between the 
transverse momentum of the two $D$ mesons $p_{t,D}$ and $p_{t, \bar D}$ 
because the single charm $p_t^2$ and $\phi$ experimental distributions agree 
well with the theory. The $\Delta \phi$ distribution shows clear evidence of 
correlations  (more details can be found in reference \cite{correlation-E791}). 

\section*{Hadronization and Particle - antiparticle asymmetries}

The production of a charm hadron can be subdivided in two steps: 
the production of a $c \bar c$ pair followed by the hadronization of
these quarks. In perturbative QCD, that describes the $c \bar c$ production, 
the $x_F$ spectra of produced charm/anticharm quarks are identical to leading
order and the effects of higher orders are very small in this respect.
Therefore any asymmetry between charm and anticharm hadrons is a simple
measure of nonperturbative effects coming from the hadronization process.\\
Particle - antiparticle asymmetries can be quantified by means of the
asymmetry parameter

\begin{equation} 
A = \frac{N - \bar N}{N + \bar N} 
\label{eq4}
\end{equation}

\noindent
where $N$ ($\bar N$) is the number of produced particles (antiparticles). 
This parameter is usually measured as a function of $x_F$ and $p_t^2$. 

Several experiments have reported an enhancement in the production rate of
charm particles having  valence quarks in common with the incident
particles, relative to charge conjugate particles which have fewer or no
common valence quarks. This effect is known as leading particle effect. 
Measurements of the asymmetry parameter A can put in evidence leading particle
effects, as well as other effects like associated production of a meson and
a baryon.\\
From the theoretical point of view, models which can account for the
presence of leading particle effects use some kind of non-perturbative
mechanism for hadronization, in addition to the perturbative production of
charm quarks. Two examples are: 

{\it String fragmentation} \cite{strings}: in this case the parton of the
hard interaction and the beam remnants are connected by a string which
reflect the confining color field. Successive breaking of the color flux
tube stretched between a cluster, when it is kinematic possible, will
create light quark-antiquark and hadrons will be produced. 

{\it Intrinsic charm model} \cite{intrinsic}: a virtual $c \bar c$ pair
pops from the sea of the beam particle. The $c \bar c$ pair coalesce with
the neighbor valence quarks due to their similar rapidity. This mechanism
favors the production of charm particles with valence quarks in common
with the beam particle at high $x_F$ and low $p_t$ region. A similar
argument can be drawn with respect to the target.\\

\begin{figure}[!ht] 
\begin{center}
\begin{tabular}{cc}
\epsfig{file=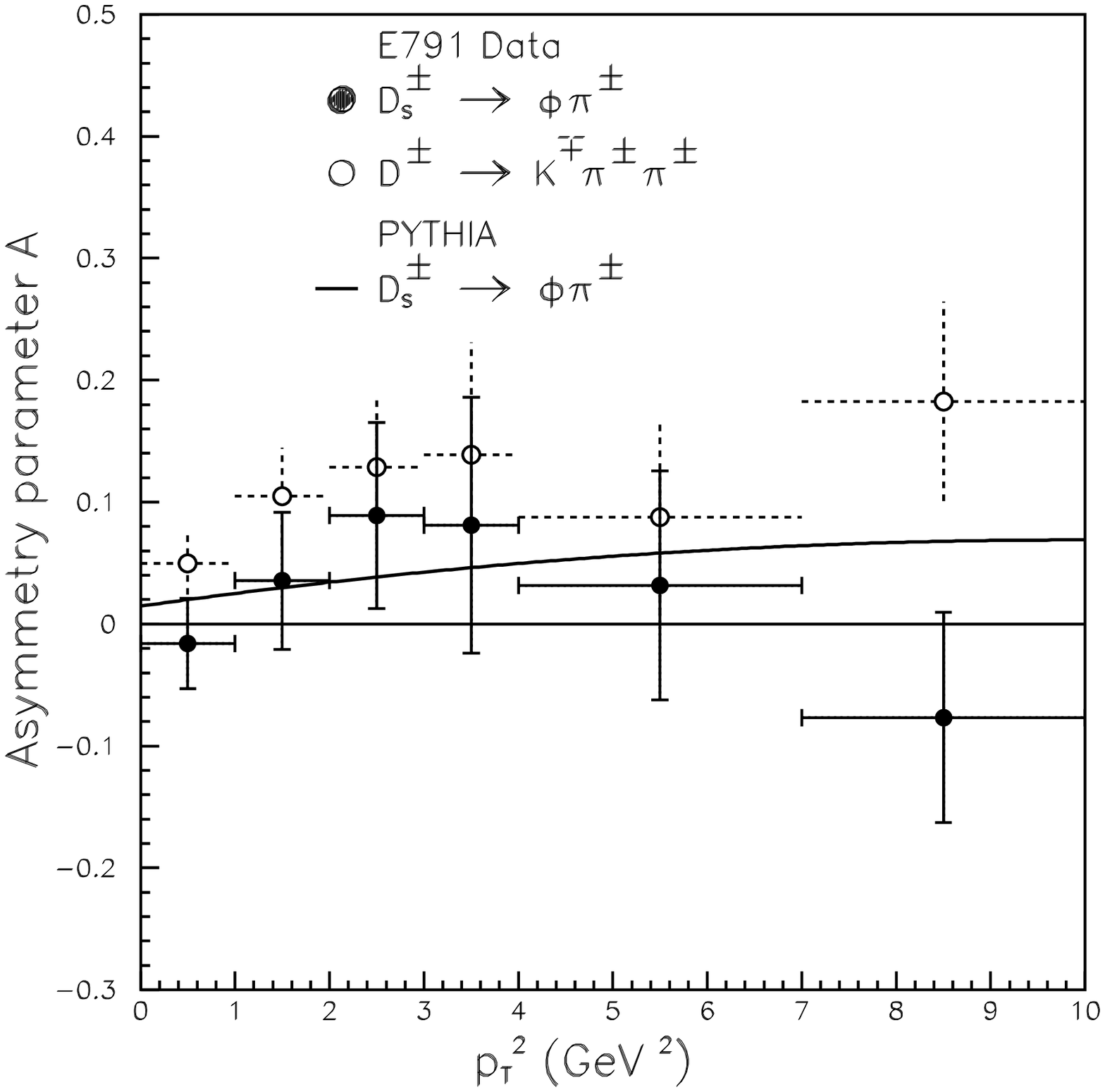,height=2.5in,width=2.5in}&
\epsfig{file=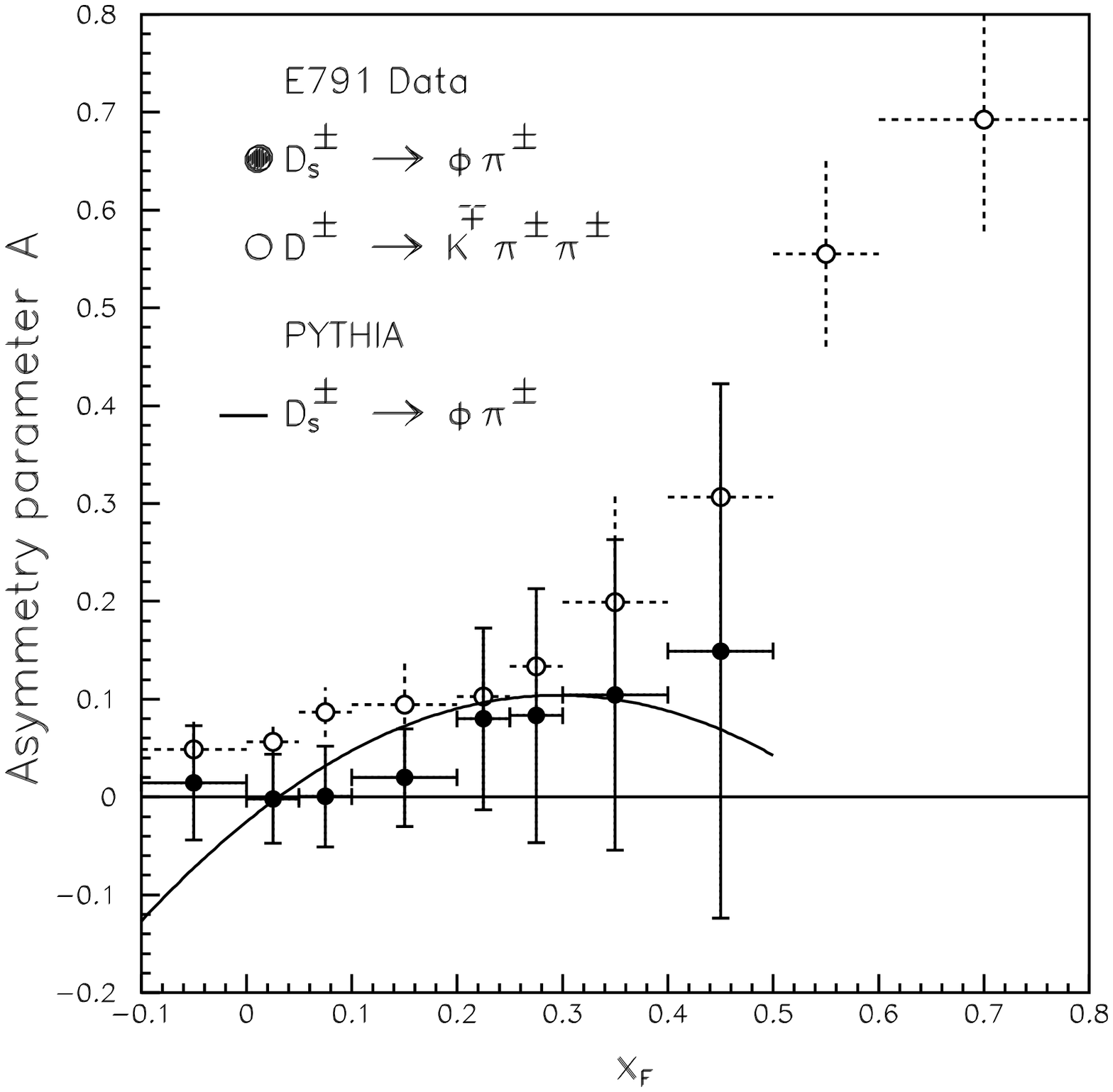,height=2.5in,width=2.5in}\\
\end{tabular}
\end{center}
\vspace{10pt}
\caption{Asymmetries of $D_s$ as a function of $x_F$ and $P_t^2$, compared 
with $D^{\pm}$.}
\label{asiDs}
\end{figure}

New results on the asymmetry parameter $A$ and evidence for leading particle
effects in both meson and baryon production are available from experiments 
E791, SELEX and FOCUS.\\
E791 has measured recently the asymmetries of $D_s^{\pm}$ mesons 
\cite{e791-ds-asi}.
Fig. \ref{asiDs} shows the $D_s^{\pm}$ asymmetry as a function of $x_F$ and 
$p_t^2$, compared with previous $D^{\pm}$ results from the same experiment 
\cite{Asim-E791}.

\begin{figure}[!ht] 
\begin{center}
\begin{tabular}{ccc}
\epsfig{file=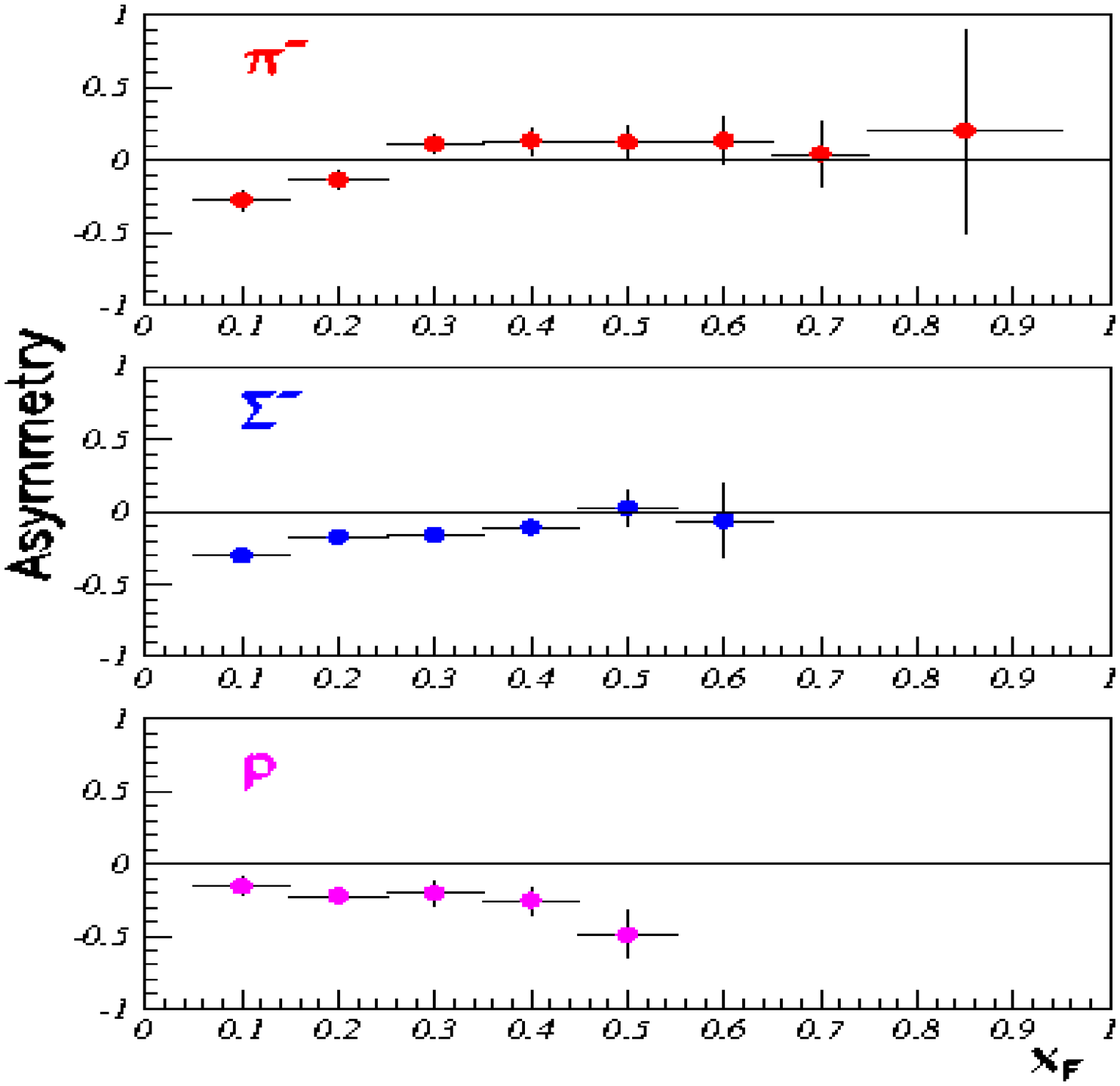,height=3.1in,width=1.5in} &
\epsfig{file=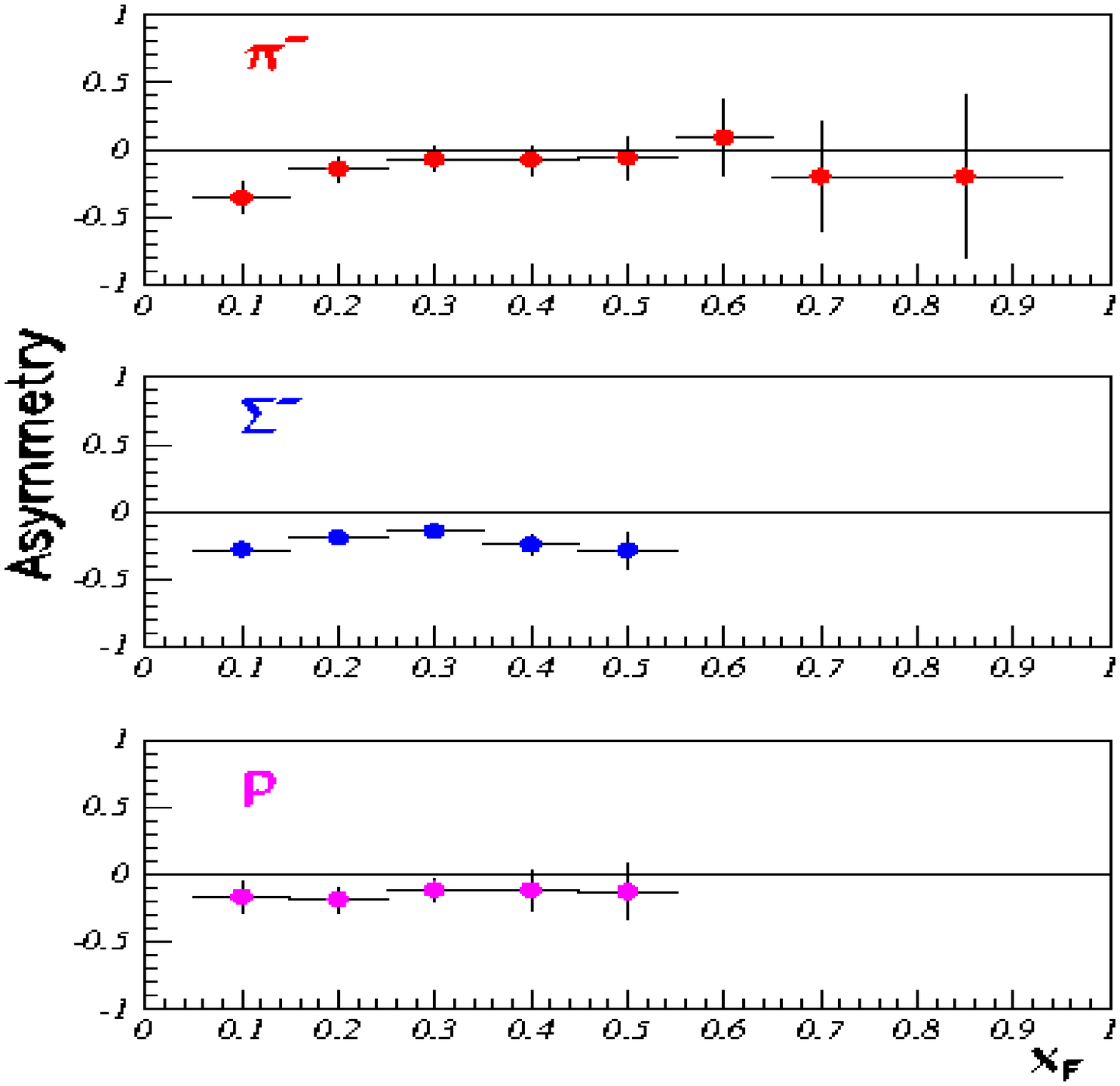,height=3.1in,width=1.5in} &
\epsfig{file=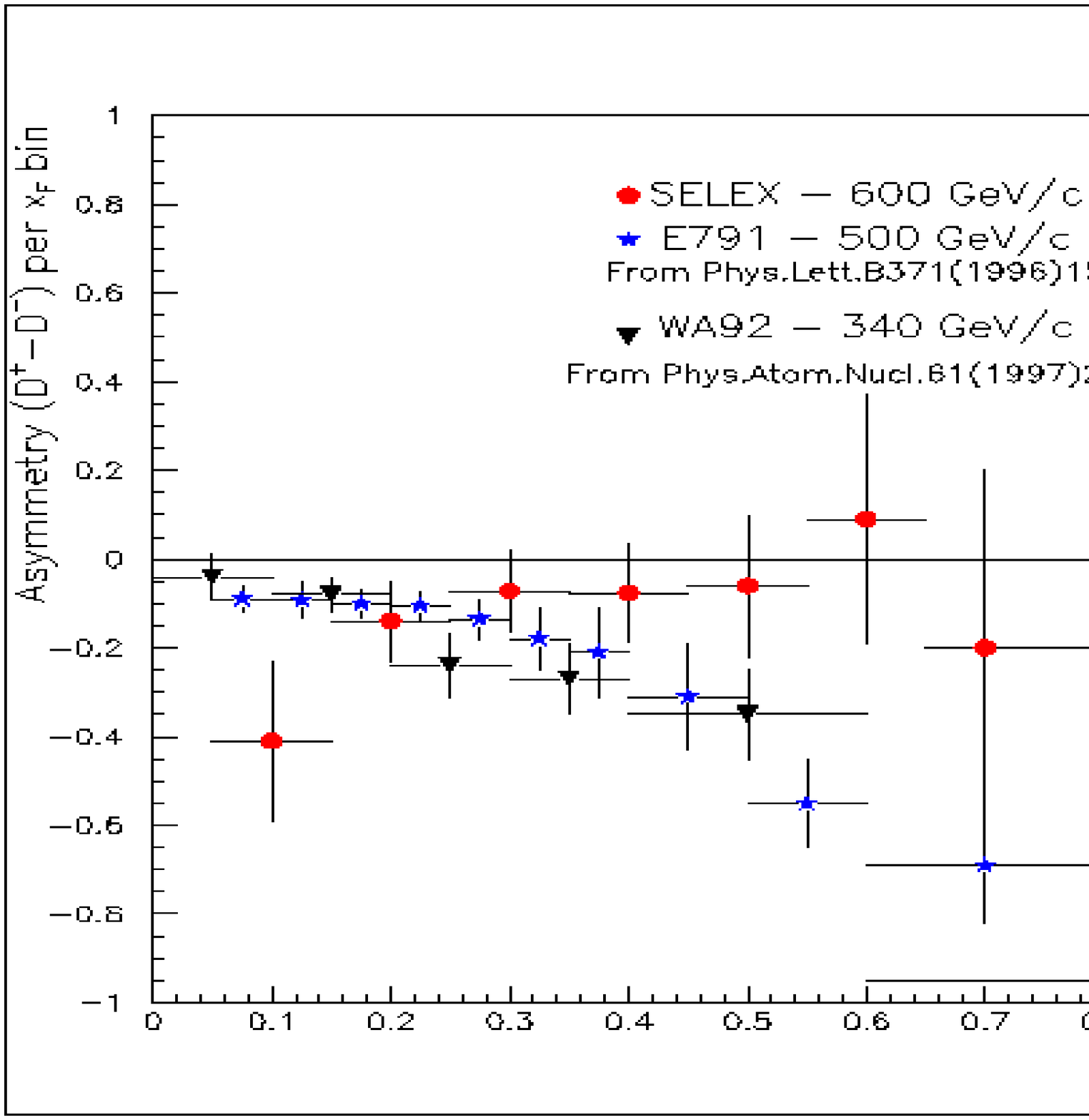,height=2.8in,width=1.5in} \\
\end{tabular}
\end{center}
\vspace{5pt}
\caption{Asymmetries of $D$ mesons from SELEX (preliminary results) and E791. 
The first and second columns are $D^{0}(\bar{D^0})$ and $D^{\pm}$ asymmetries 
from SELEX. The right figure shows the comparison of $D^{\pm}$ 
asymmetries from  SELEX and E791.} 
\label{asiD00}
\end{figure}

\noindent
Preliminary results of $D^{0}(\bar{D^0})$  as well as 
$D^{\pm}$ asymmetries as a function of $x_F$ presented by SELEX \cite{Lori}  
are shown in Fig. \ref{asiD00} (left), for different incident beam particles 
($\pi^-$, $\Sigma^-$, $p$). The $\pi^-$ data is compared to $D^{\pm}$ 
asymmetries from 
E791. The asymmetry at $x_F \: > \:0.4$ does not rise steeply with $x_F$ as 
previously reported by E791, Fig. \ref{asiD00} (right).

\begin{figure}[!ht] 
\begin{center}
\begin{tabular}{ccc}
\epsfig{file=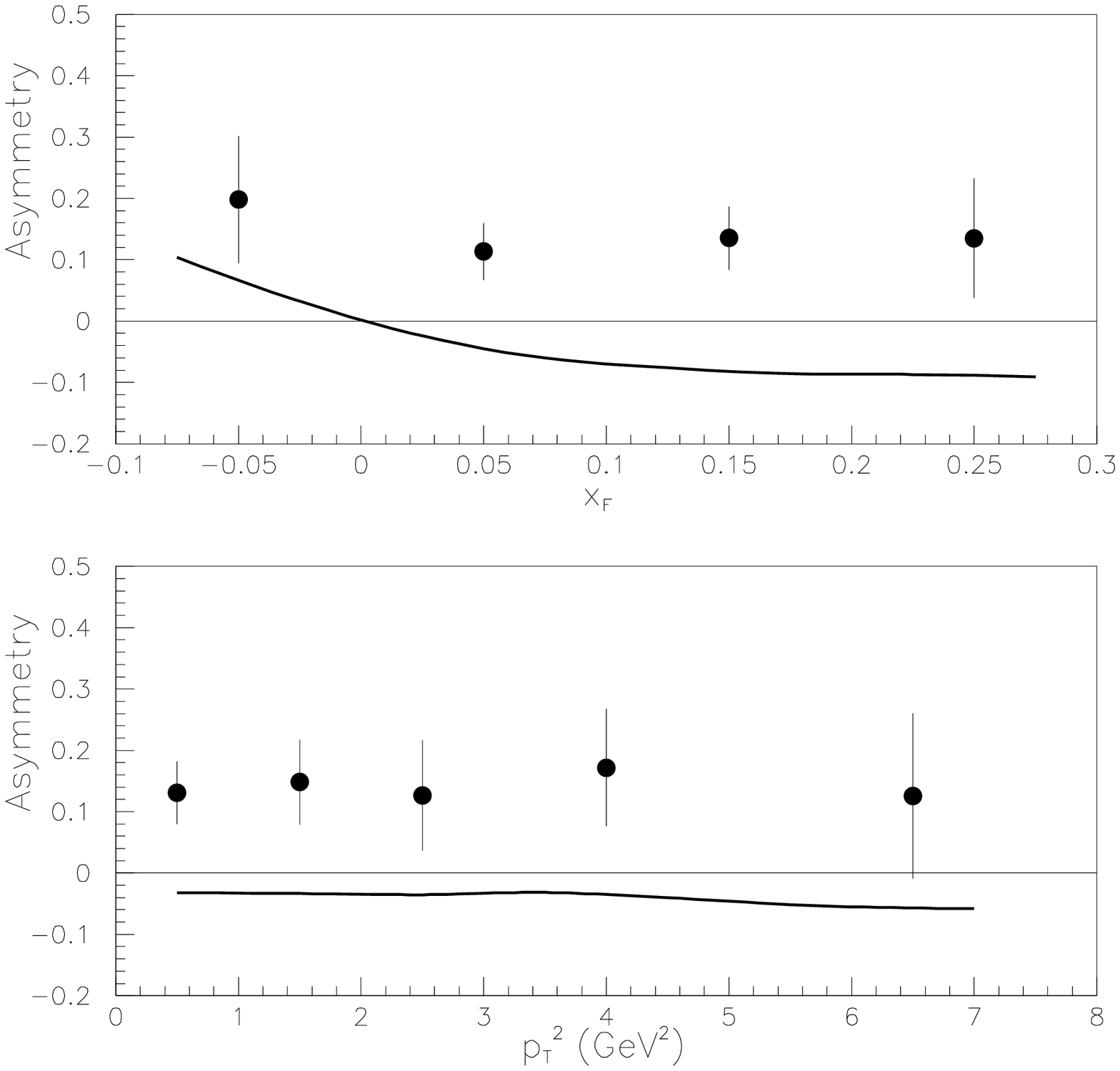,height=2.8in,width=2.in}& &
\epsfig{file=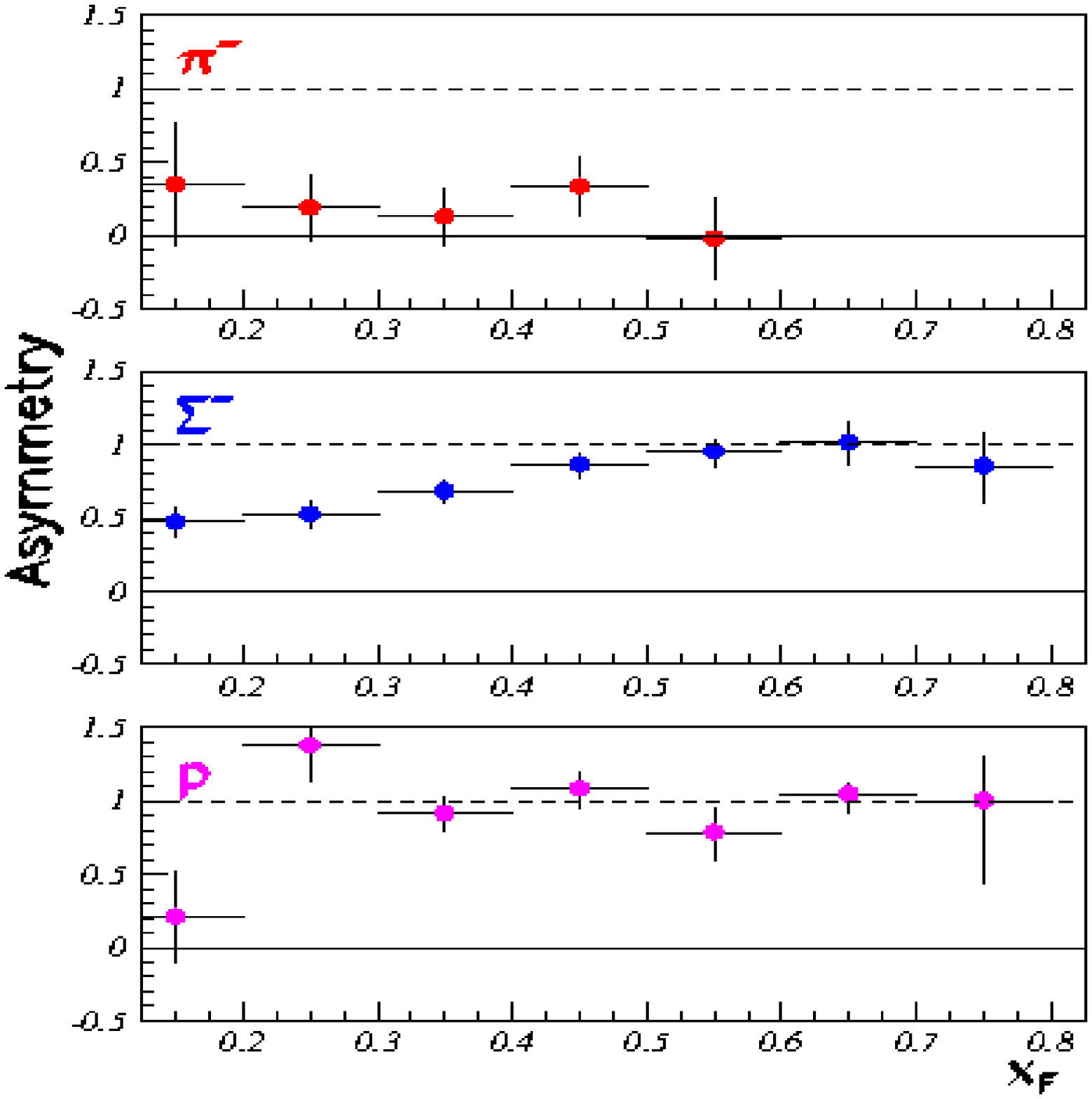,height=3.in,width=2.1in}\\
\end{tabular}
\end{center}
\vspace{3pt}
\caption{$\Lambda_c$ asymmetries as function of $p_t^2$ and $x_F$ from  
E791 (left). $\Lambda_c$ asymmetries (preliminary results) as function of 
$x_F$ for different beams ($\pi^-, \: \Sigma^-, \: p$) from SELEX (right).}
\label{aLselex791}
\end{figure}

\noindent
For  baryons there are also  preliminary results for the asymmetry parameter. 
Fig.\ref{aLselex791}
(left) shows the E791 results for the $\Lambda_c^+$ asymmetries as function 
of $x_F$ and $p_t^2$, compared with predictions from Pythia/Jetset 
(full lines) \cite{magnin}. The results show a uniform positive asymmetry of 
$12.7 \pm 3.4\%$ over the 
studied  kinematical range but do not exclude a rise in the $x_F \:< \:0$ 
region as predicted by Pythia/Jetset. For $x_F>0$ the observed asymmetry does 
not agree with Pythia/Jetset predictions.\\
SELEX has also measured the $\Lambda_C^+$ asymmetry as a function of $x_F$ 
for different incident beam particles ($\pi^-, \: \Sigma^-, \: p$)  Fig. 
\ref{aLselex791} (right) \cite{Lori}. The asymmetry is clearly larger for 
the baryon beams than for the $\pi$ beam. For the protons the only region in 
which there is $\bar \Lambda_c$ production is at very small $x_F$. Their 
preliminary results for the $\pi^-$ beam are compatible with those from E791, 
Fig. \ref{aLselex791} (left).

\begin{figure}[!ht] 
\begin{center}
\begin{tabular}{ccc}
\epsfig{file=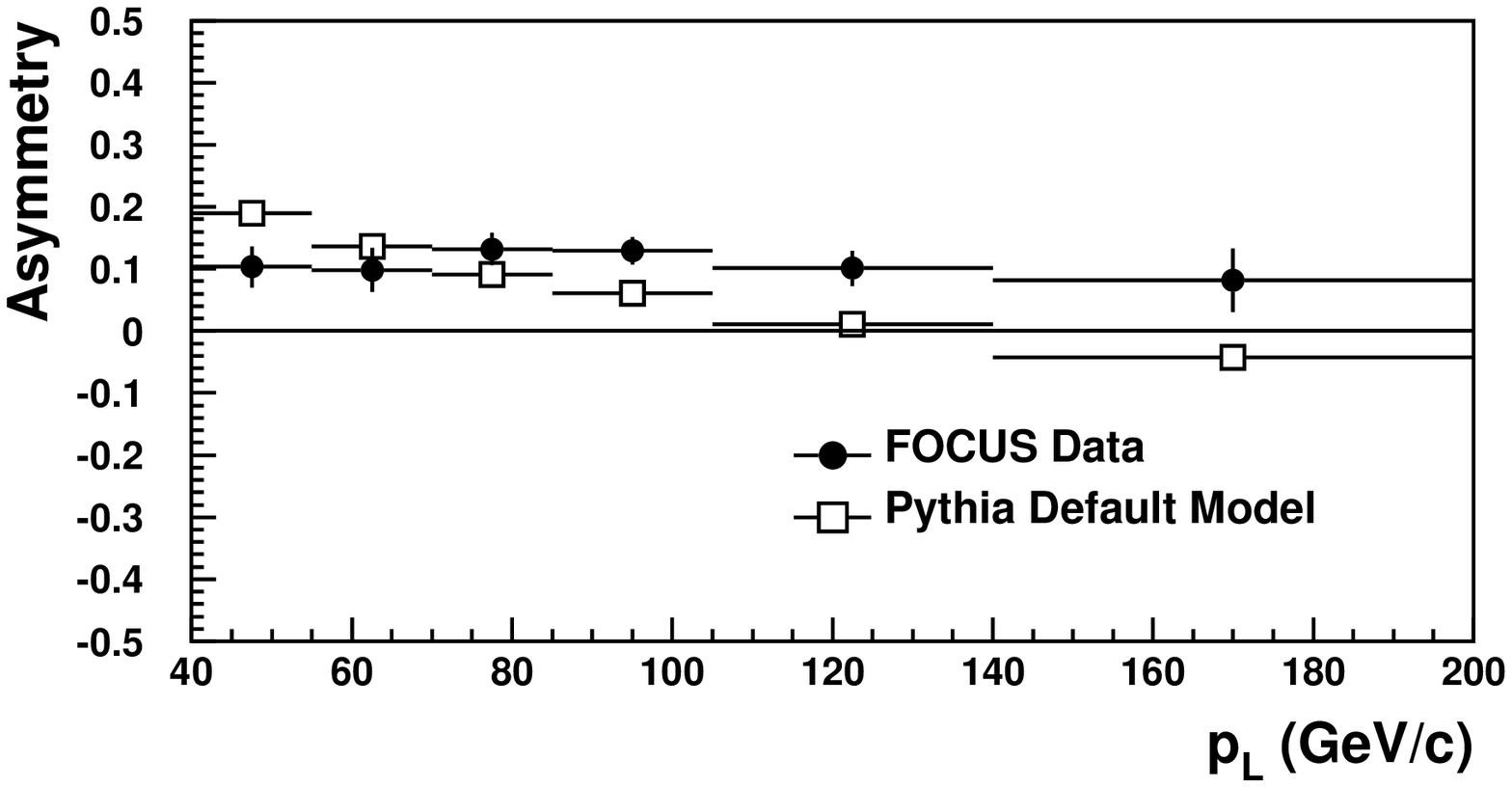,height=1.8in,width=1.8in}&
\epsfig{file=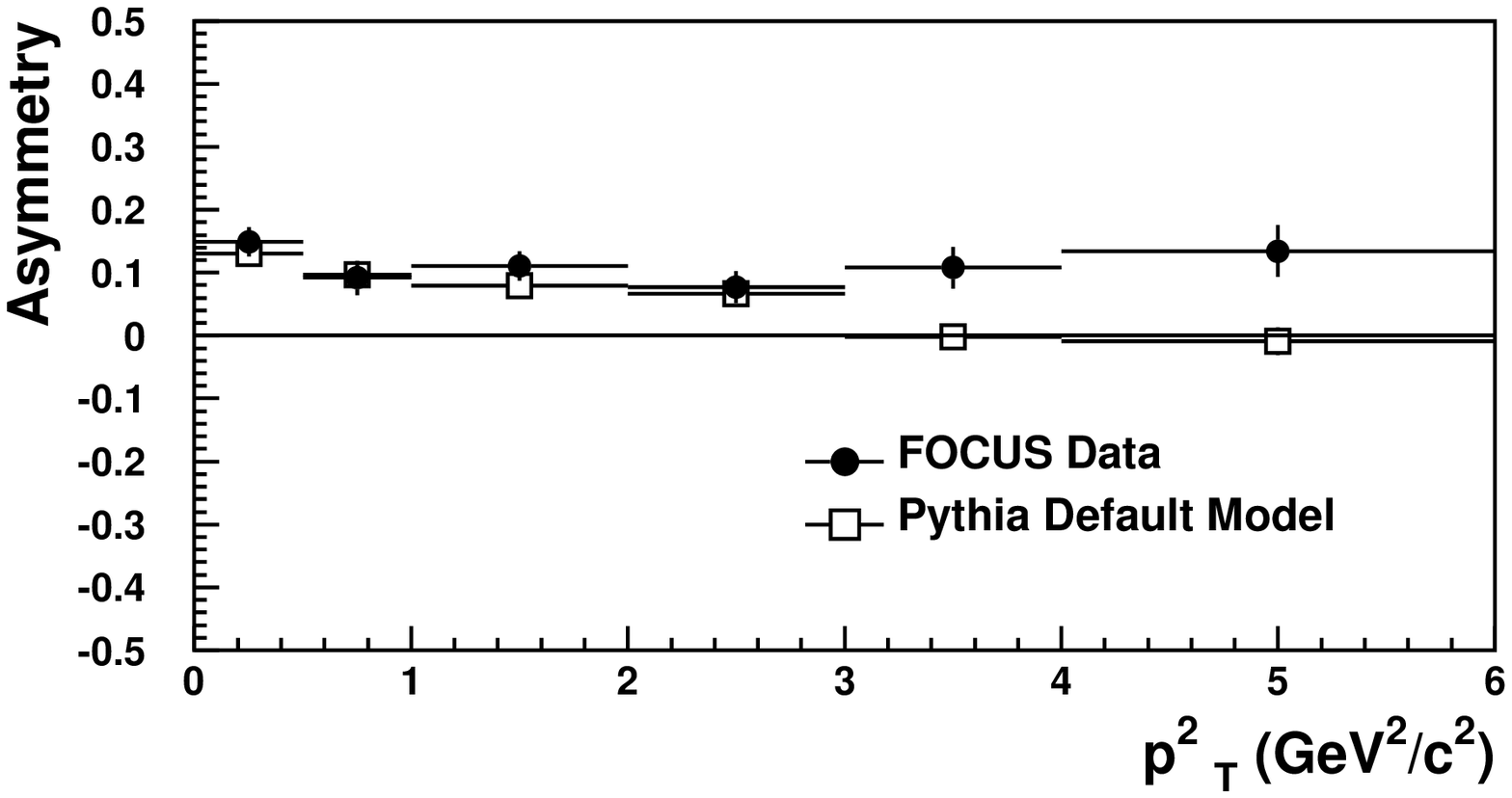,height=1.8in,width=1.8in}&
\epsfig{file=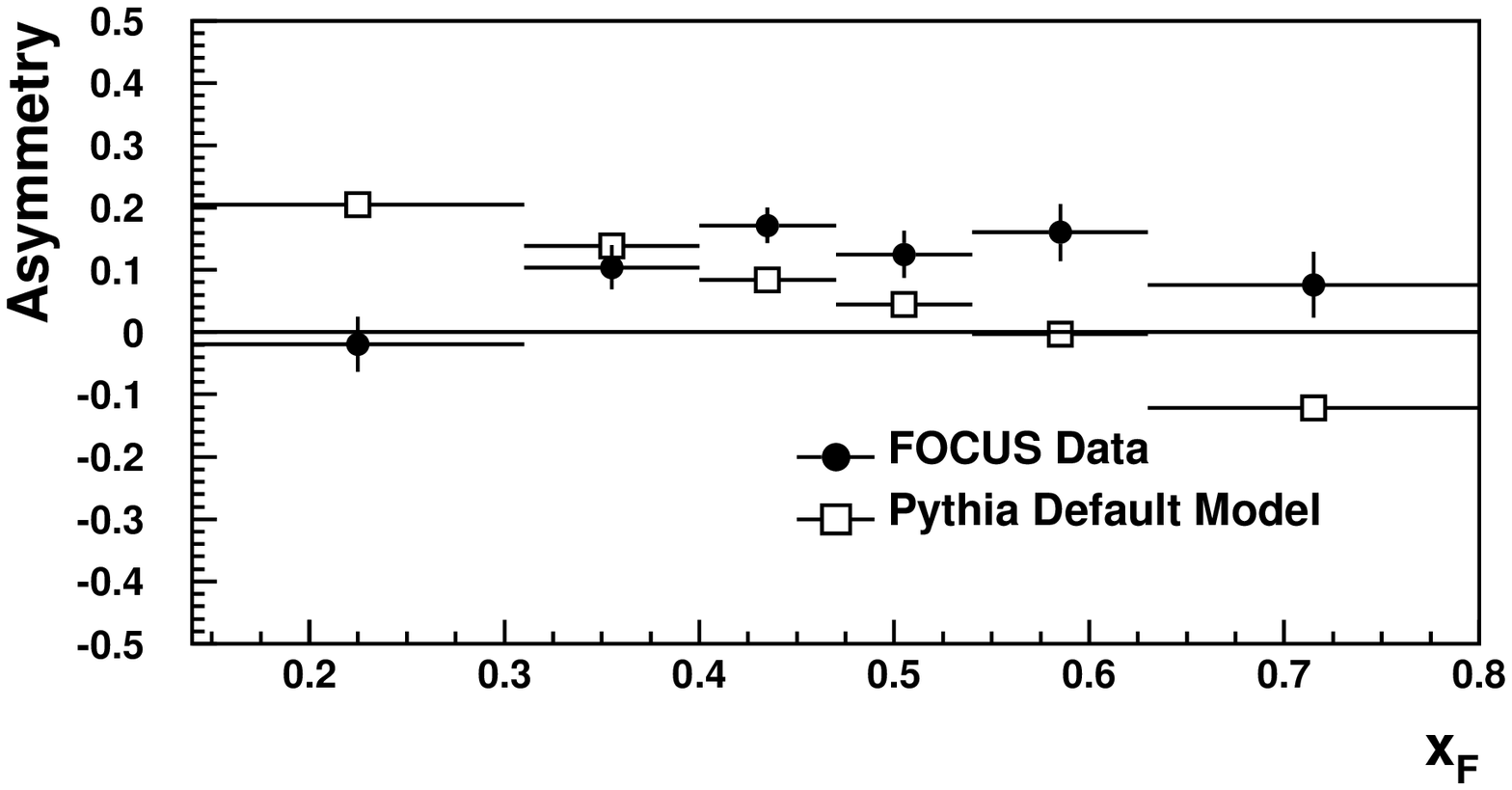,height=1.8in,width=1.8in}\\
\end{tabular}
\end{center}
\vspace{10pt}
\caption{$\Lambda_c^+$ asymmetry from FOCUS as a function of 
$p_l, \: p_t^2$ and $x_F$ (preliminary results), compared with Pythia/Jetset 
predictions} 
\label{fig8}
\end{figure}

\begin{figure}[!ht] 
\begin{center}
\begin{tabular}{cc}
\epsfig{file=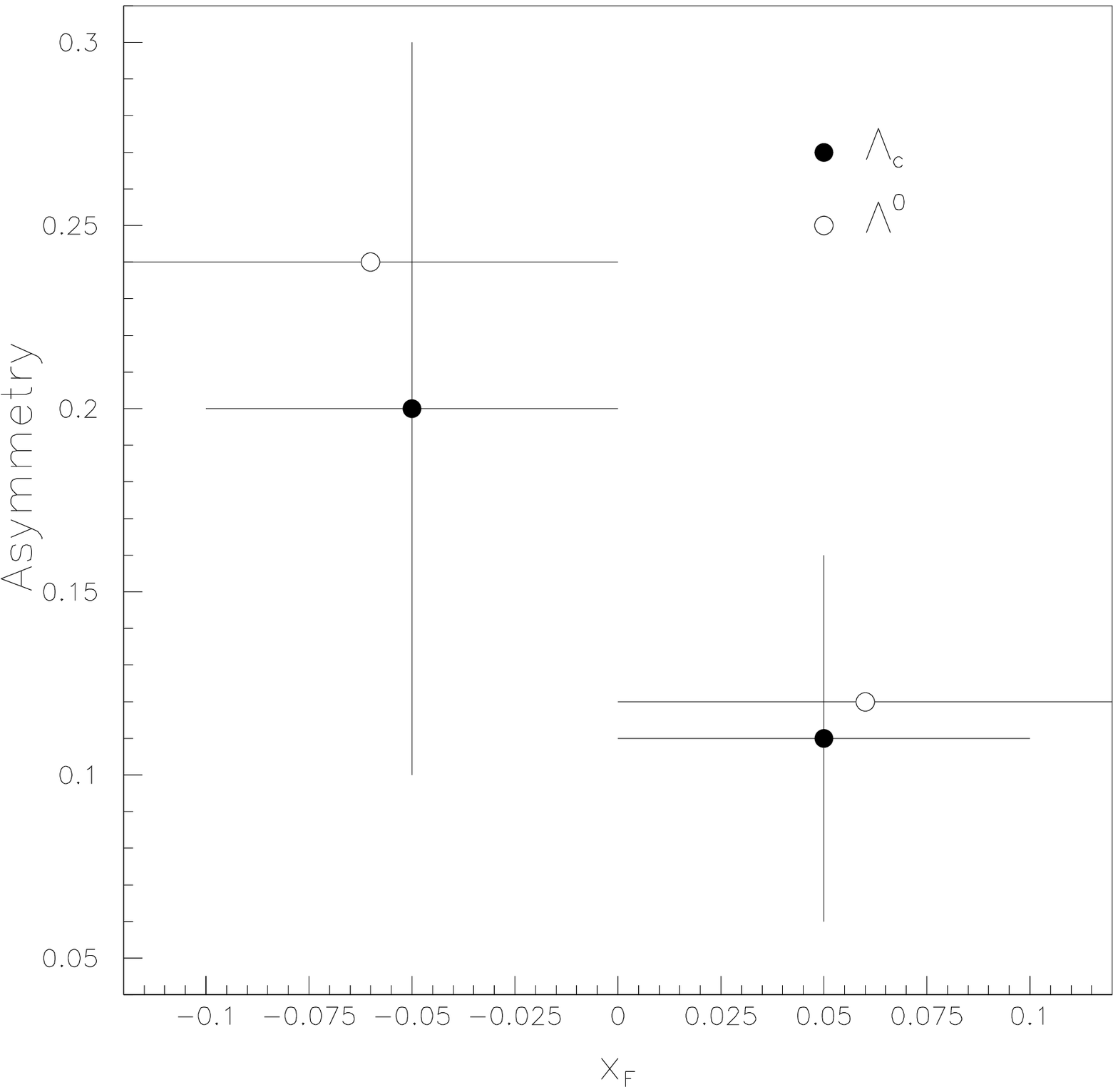,height=2.1in,width=2.in} &
\epsfig{file=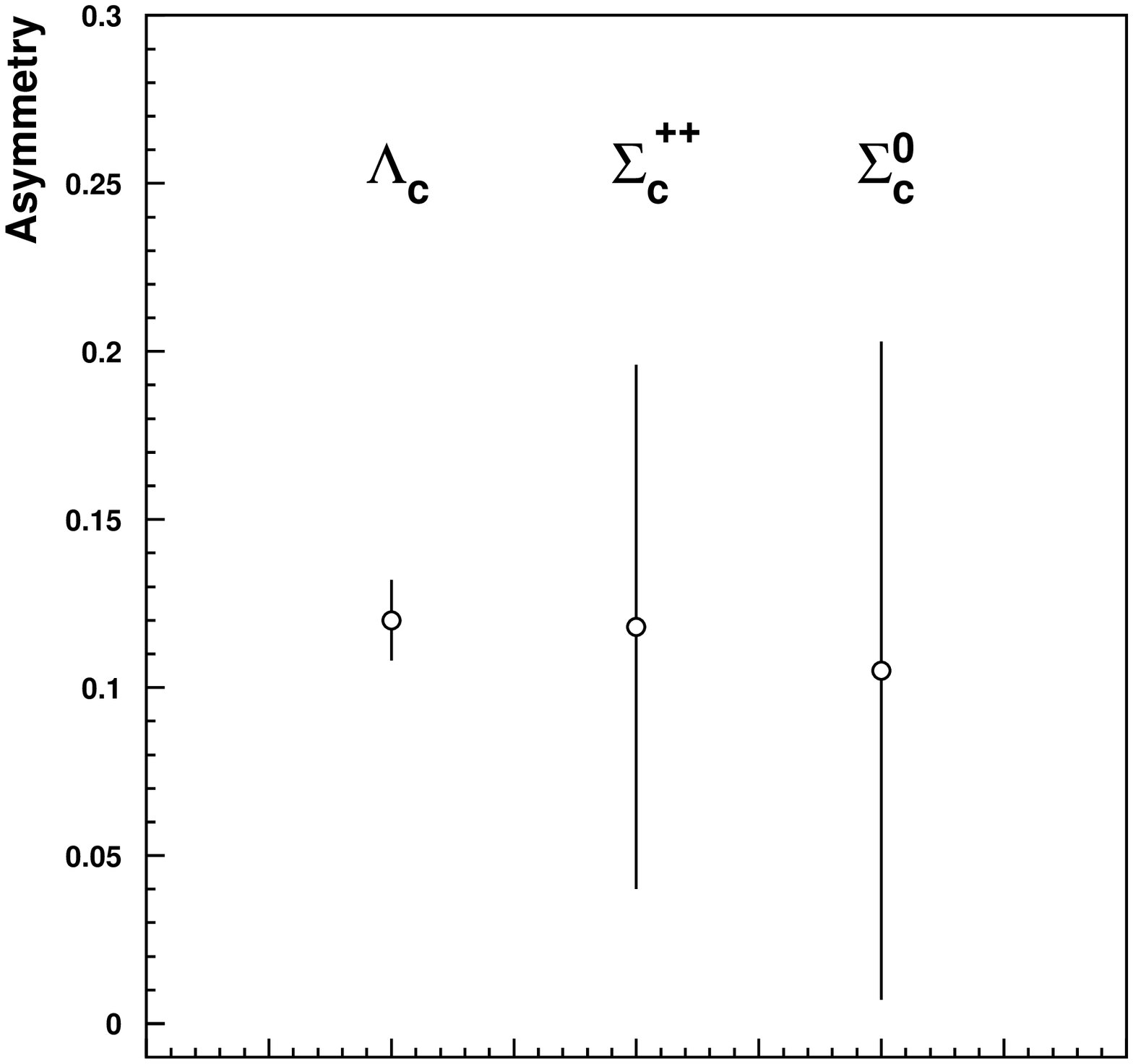,height=2.in,width=2.in}\\
\end{tabular}
\end{center}
\vspace{10pt}
\caption{Comparison between $\Lambda^0$ and $\Lambda_c$ asymmetries from E791 
(left). $\Lambda_c$,  $\Sigma^{++}$ and $\Sigma_c^0$ total 
asimmetries (preliminary) from FOCUS (right).}
\label{fn3}
\end{figure}

\noindent
FOCUS has also preliminary results on baryon asymmetries. In 
Fig.\ref{fig8} we see the $\Lambda_c$ asymmetry as functions of $p_l$, $p_t^2$
and $x_F$  obtained from a
sample of about 16,000 $\Lambda_c 's$, compared with Pythia/Jetset 
Predictions. As FOCUS has a photon beam, no leading
particle effect is expected in the $x_F >0$ region. In this case the positive
asymmetry observed in all the $x_F$ range can be an indication of charm
baryon and charm meson associated production, favouring a positive
asymmetry.\\
The high statistic $\Lambda_c$ sample from FOCUS allowed to obtain about 600
$\Sigma_c \rightarrow \Lambda_c \pi$.
It is interesting to compare the asymmetry for charm particles with different 
light quark content. We present in Fig.\ref{fn3} (right) 
very  preliminary results from FOCUS comparing the $\Sigma_c^{++}(uuc)$ and
$\Sigma_c^{0}(ddc)$ to the $\Lambda_c^{+}(udc)$ total asymmetry.\\
In Fig. \ref{fn3} (left) we show the comparison between the $\Lambda^0$ 
and the $\Lambda_c$ asymmetries as a function of $x_F$ from E791. Their 
similarity suggests that the $ud$ diquark shared by the produced $\Lambda^0$ 
($\Lambda_c^+$)  and 
nucleons in the target should play an important role in the measured 
asymmetry in the $x_F < 0$ region. However, one expects that $\Lambda_c$ 
asymmetry grows more slowly than the $\Lambda^0$ asymmetry due to its 
higher mass.\\

\begin{figure}[!ht] 
\centerline{\epsfig{file=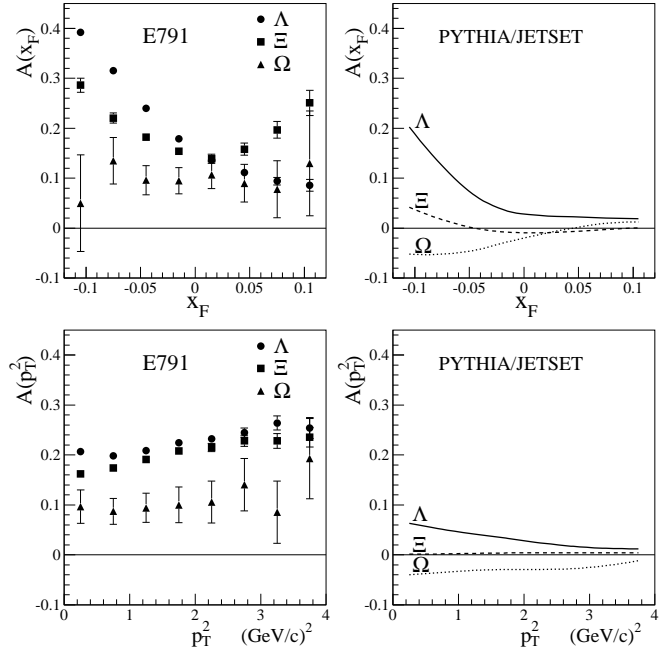,height=3.5in,width=3.5in}}
\vspace{10pt}
\caption{Hyperon asymmetries as a function of $x_F$ (top) and 
$p_T^2$ (bottom) from E791 (preliminary results). The asymmetry for $x_F$ 
($p_T^2$) is integrated over all  the $p_T^2$ ($x_F$) range of the data set. 
The right column  show the predictions of Pythia/Jetset both as a function of 
$x_F$ and $p_T^2$  (preliminary results).}
\label{fn1}
\end{figure}

\noindent
Leading particle effects were also seen by E791 in hyperon production. 
Preliminary results on $\Lambda$, $\Xi$ and $\Omega$ asymmetries 
as a function of $x_F$ and $p_t^2$ are shown in Fig.\ref{fn1} in comparison 
with  predictions from
Pythia/Jetset. The range of $x_F$ covered allowed the first simultaneous
study of the asymmetry in both the negative and positive $x_F$ regions. We
can clearly see leading particle effects associated with the target or
with the beam particles which qualitatively agree with expectations from
recombination models (see table \ref{hyperones}) \cite{Ahyper}. It is 
interesting to 
observe, as expected, the crossover of the $\Xi$ asymmetry with respect to 
the $\Lambda$ asymmetry at $x_F \: \simeq \: 0$. The positive asymmetry 
measured in regions 
$x_F > 0$ for the $\Lambda (udc)$  and for the $\Omega (sss)$ suggest the 
associated  production of a hyperon
and a kaon due to the higher energy threshold imposed by baryon number
conservation for the production of an anti-hyperon. Pythia/Jetset does not 
reproduce the data. 

\begin{table}
\caption{Hyperon asymmetries predictions}
\label{hyperones}
\begin{tabular}{lccc}
  &  $x_f < 0$  &  $x_f > 0$  \\
  &  target(uud or ddu) & beam $\pi^- \: (\bar u d)$ \\
\tableline
$\Lambda^0 (uds)$  &     double leading   &        leading \\
$\overline \Lambda^0 (\bar u \bar d \bar s)$ &  non leading &  leading \\
$\Xi^- (dss)$   &           leading    &             leading \\
$\overline \Xi^+ (\bar d \bar s \bar s)$  &  non leading  & non leading \\
$\Omega^- (sss)$   &           non leading   &       non leading  \\
$\overline \Omega^+ (\bar s \bar s \bar s)$&  non leading & non leading \\
\tableline
\tableline
Recomb. Models\tablenote{Presented in Second Latin American Symposium 
\cite{Ahyper}}: & 
$ \: A_{\Lambda} >  A_{\Xi}  >  A_{\Omega} \:\:$& 
$ \:\: A_{\Xi} >  A_{\Lambda} \sim A_{\Omega}$  \\ 
\end{tabular}
\end{table}

\section*{Double Cabibbo suppressed decays}

\begin{figure}[!ht] 
\begin{center}
\begin{tabular}{cccc}
\epsfig{file=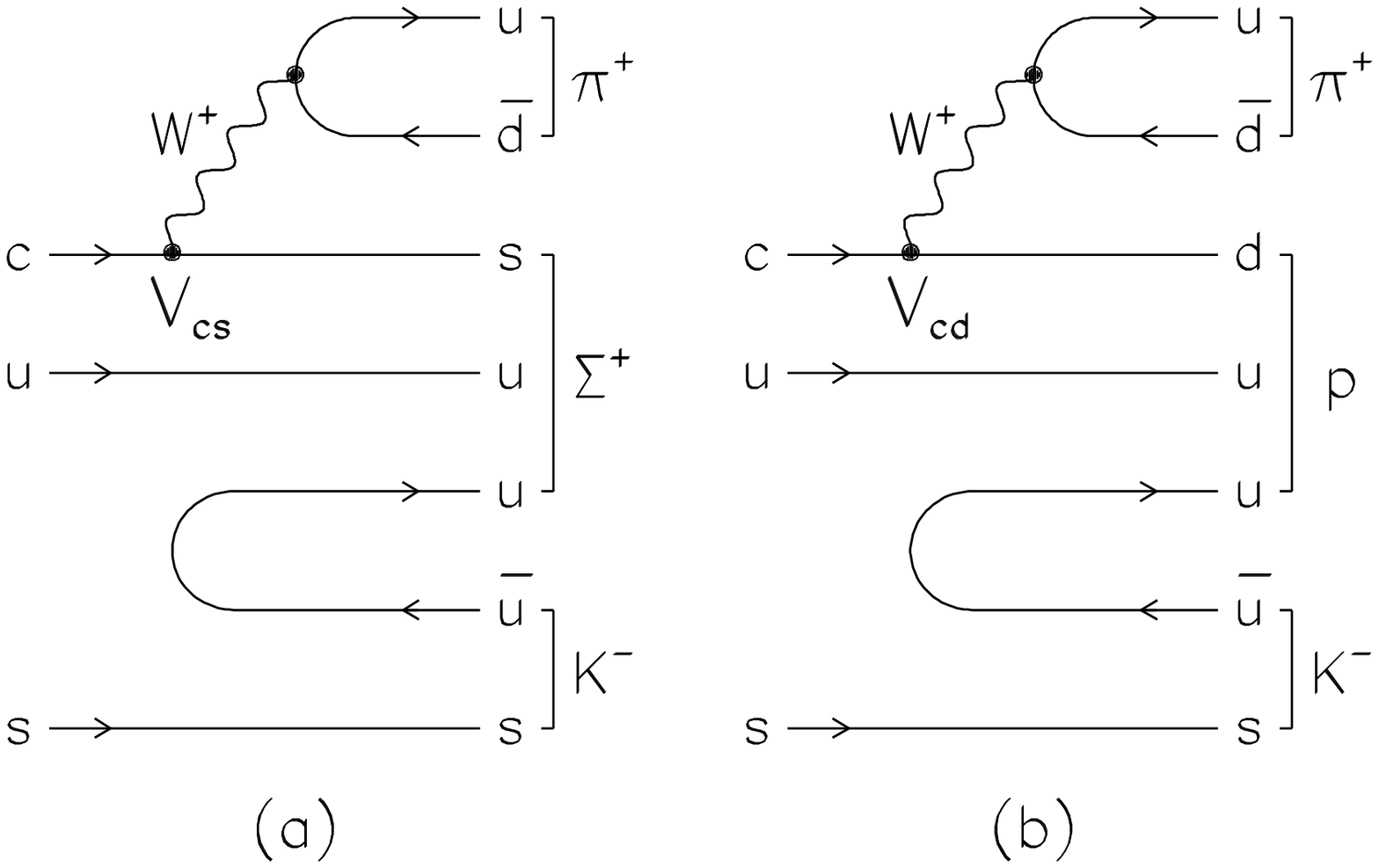,height=2.5in,width=2.5in} & & &
\epsfig{file=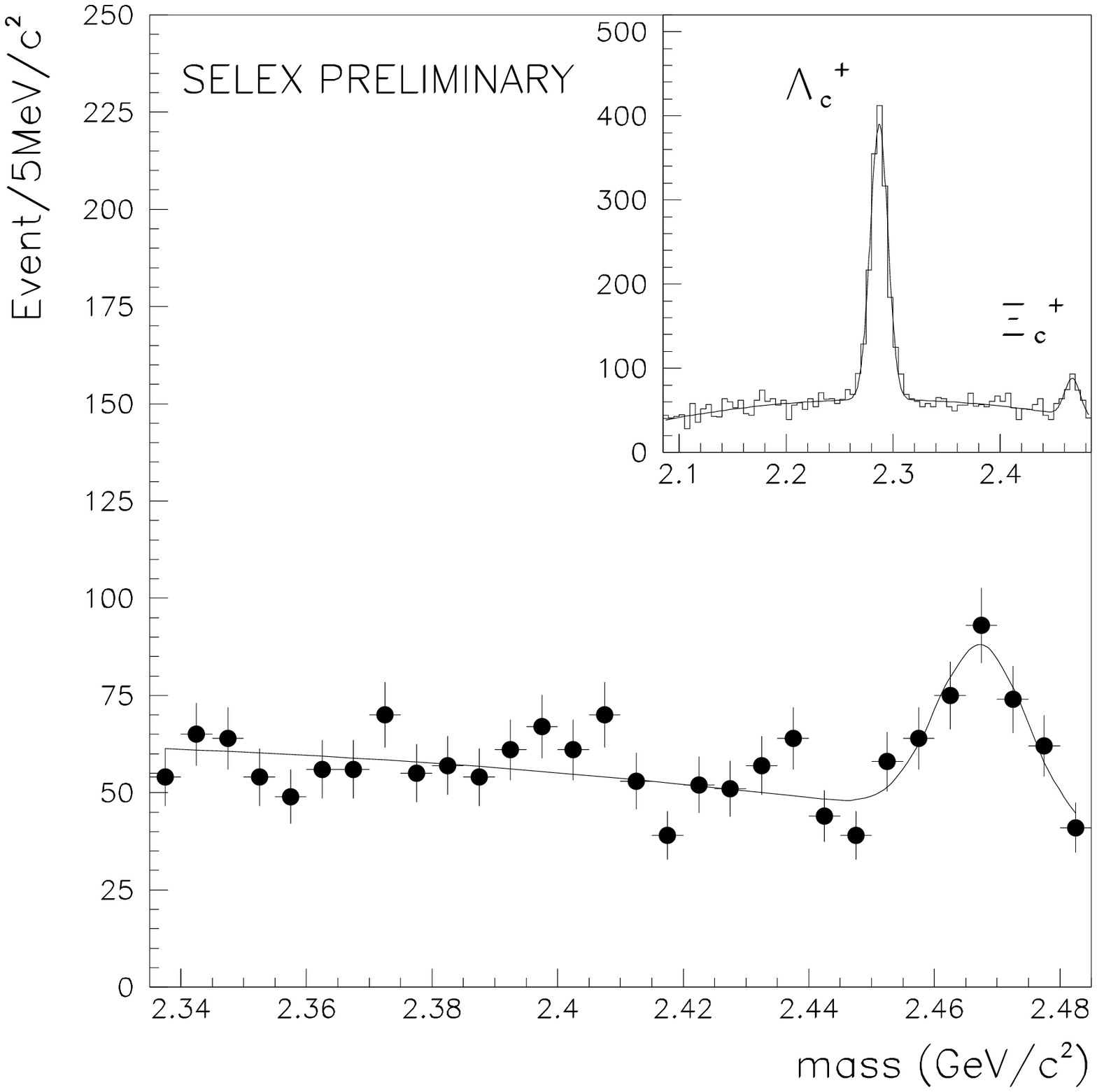,height=3.in,width=2.5in}\\
\end{tabular}
\end{center}
\vspace{10pt}
\caption{Simple spectator diagrams (a) and (b) and preliminary signal of 
$\Xi_c^+ \rightarrow p K^- \pi^+$  from SELEX.}
\label{fig11}
\end{figure}

The Cabibbo suppressed charm decays can provide useful insights into the
weak interaction mechanism for nonleptonic decays. The $D^+
\rightarrow K^+ \pi^- \pi^+ $ signal obtained from 100 $\%$ of FOCUS data
set consist of $\sim$ 300 events and is at least a factor of five larger
than two previous observations by E687 and E791 ( E791 observed $59 \pm 13$ 
events \cite{e791-dcsd}). 
The preliminary branching
ratio relative to $K^- \pi^+ \pi^-$ is ($0.72 \pm 0.09$) $\%$, completely
consistent with the world average of ($0.68 \pm 0.15$)$\%$ and the E791 
values of
($0.77 \pm 0.17 \pm 0.08$)$\%$. We note that this is
$\sim$ $3 tan^4(\theta_c)$, which is roughly the ratio of the $D^+/D^0$
lifetime, indicating that the destructive Pauli interference present in
the Cabibbo Favored $D^+$ decay is absent in the doubly Cabibbo suppressed
(DCS) mode. \\
$D^+ \rightarrow K^- K^+ K^+$ is an interesting DCS decay, which cannot 
even occur through a spectator diagram. FOCUS has the first observation 
of this mode, and reports a preliminary result for the Branching Ratio 
relative to $K^- \pi^+ \pi^+$ of ($0.14\pm 0.02$) $\% $. 
Several groups have reported observations of a $D^+ \rightarrow \phi K^+$ 
signal, however FOCUS did not find evidence for such decay \cite{brian-dcsd}.\\
SELEX announced the first observation of a Cabibbo suppressed decay of a 
charm baryon through the decay $\Xi_c^+ \rightarrow p K^- \pi^+$ 
\cite{printSelex}. Fig. \ref{fig11} shows the signal of $157 \pm 22$ 
events reported by SELEX and simple spectator 
diagrams with external $W$ emission for $\Xi_c^+$ decaying into a Cabibbo 
allowed and into a single Cabibbo suppressed (SCS) mode. The other Cabibbo 
allowed $\Xi^-$ mode interchanges $s$ and $d$ quarks lines and produces a 
$d \bar d$ pair from the vacuum instead of a $d \bar u$ pair. 
FOCUS has also observed the same SCS decay, reporting a signal of $86 \pm 
21$ events from about 70 $\%$ of their data.\\

E791 has published \cite{plb423-98-185} results on the singly Cabibbo 
suppressed 
decay, $D^0 \rightarrow K^- K^+ \pi^- \pi^+$. A coherent amplitude 
analysis of the resonant substructure was used to extract decay 
fractions. Significant phase angles among different modes indicate very 
strong interference. The measured branching fractions relative to $D^0 
\rightarrow K^- \pi^+ \pi^- \pi^+$ are presented in the table \ref{br791}.

\begin{table}
\caption{Branching fractions relative to 
$D^0 \rightarrow K^- \pi^+ \pi^- \pi^+$ from E791.}
\label{br791}
\begin{tabular}{lcc}
\multicolumn{1}{c}{Mode}&\multicolumn{1}{c}{Branching Fraction}\\
\tableline
$\phi \rho ^0$               &  ($2.0 \pm 0.9 \pm 0.8$) $\%$ \\
$\phi \pi^+ \pi^-$           &  ($0.9 \pm 0.4 \pm 0.5$) $\%$ \\
$\overline{K^{*0}} K^{*0}$   &  $ < 2.0 \: \% \: (90 \: \% \:CL)$ \\ 
$\overline{K^{*0}} K^+ \pi^-$ + $\overline{K^{*0}} K^- \pi^+$  &
$< 2.0 \: \% \:\: (90 \: \% \: CL)$ \\
\end{tabular}
\end{table}


\subsection*{Hadronic charm decays, Dalitz plot Analysis}

With the advent of high statistics experiments, charm meson decay have
become a new way to study light meson spectroscopy. The amplitude
analysis performed on Dalitz plots gives insight into the decay
dynamics, providing direct information about intermediate resonances
and relative decay fractions, and allowing to study final state
interactions coming from the interference of the amplitudes describing
competing resonant channels. 

E791 has preliminary results on the decay of $D^+$ and $D_s^+$ mesons
in three pions. A clear signal with $1240\pm 51$ $D^+$ and $858\pm 49$
$D_s^+$ was obtained after applying selection criteria aimed at identifying 
a clearly
separated $3\pi$ vertex and after carefully estimating the backgrounds
coming from possible reflections and three pion combinations.  The
branching ratios were normalized to $D^+ \rightarrow K^- \pi^+ \pi^+$
($34,790\pm 232$ events) and to $D_s^+ \rightarrow \phi \pi^+$
($1038\pm 44$ events) respectively. Efficiencies were obtained from a
full Monte Carlo simulation. The branching ratio of the $D^+
\rightarrow \pi^+ \pi^- \pi^+$ relative to $D^+ \rightarrow K^- \pi^+
\pi^+$ obtained was $0.0329 \pm 0.0015^{+0.0016}_{-0.0026}$. Similarly
the branching ratio of the $D_s^+ \rightarrow \pi^+ \pi^- \pi^-$
relative to $D_s^+ \rightarrow \phi \pi^+$ was 
$0.247 \pm 0.028^{+0.019}_{-0.012}$. 

\subsubsection*{$D_s^+$ Dalitz plot results from E791}
Among the advantages of using charm meson decays to study light I=J=0
states is the fact that, unlike hadron-hadron scattering, in the decays
of $D$ mesons the initial state is always $J^P = 0^-$, limiting the
number of possible final states. The decay $D_s^+ \rightarrow \pi^-
\pi^+ \pi^+$ is Cabibbo-favored without a strange meson in the final
state. It can proceed via spectator amplitudes producing intermediate
resonant states with hidden strangeness like the $f_0 (980)$ or it can
 proceed via W-annihilation amplitudes producing intermediate resonant
states with no strangeness. The decays like $D_s^+ \rightarrow \rho^0
\pi^+$ and the non-resonant $D_s^+ \rightarrow \pi^- \pi^+ \pi^+$ would
proceed via $W$-annihilation mechanism. It would also be responsible
for the decay $D_s^+ \rightarrow f_0(1370) \pi^+$, if the $f_0(1370)$
resonance is at least partially a $u \bar u$ + $d \bar d$ state as
predicted by the simple quark model.  The scale of the $W$-annihilation
compared to the $W$-radiation amplitude would be indicated by the
relative contribution of these channels to the $\pi^- \pi^+ \pi^+$
final state \cite{albertoreis}. 

\begin{figure}
\begin{center}
\begin{tabular}{cccc}
\epsfig{file=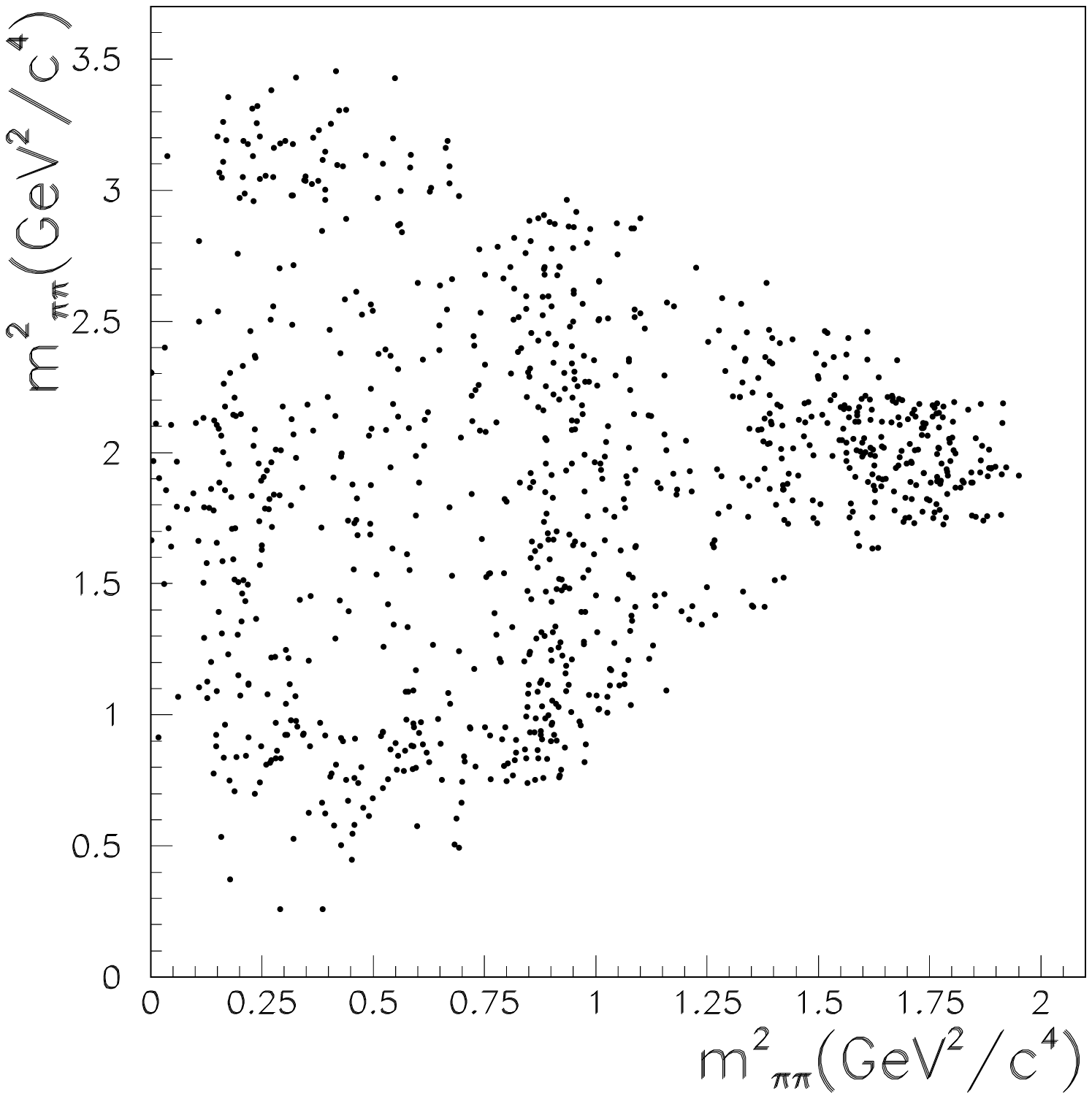,height=1.5in,width=2.in} & &
\epsfig{file=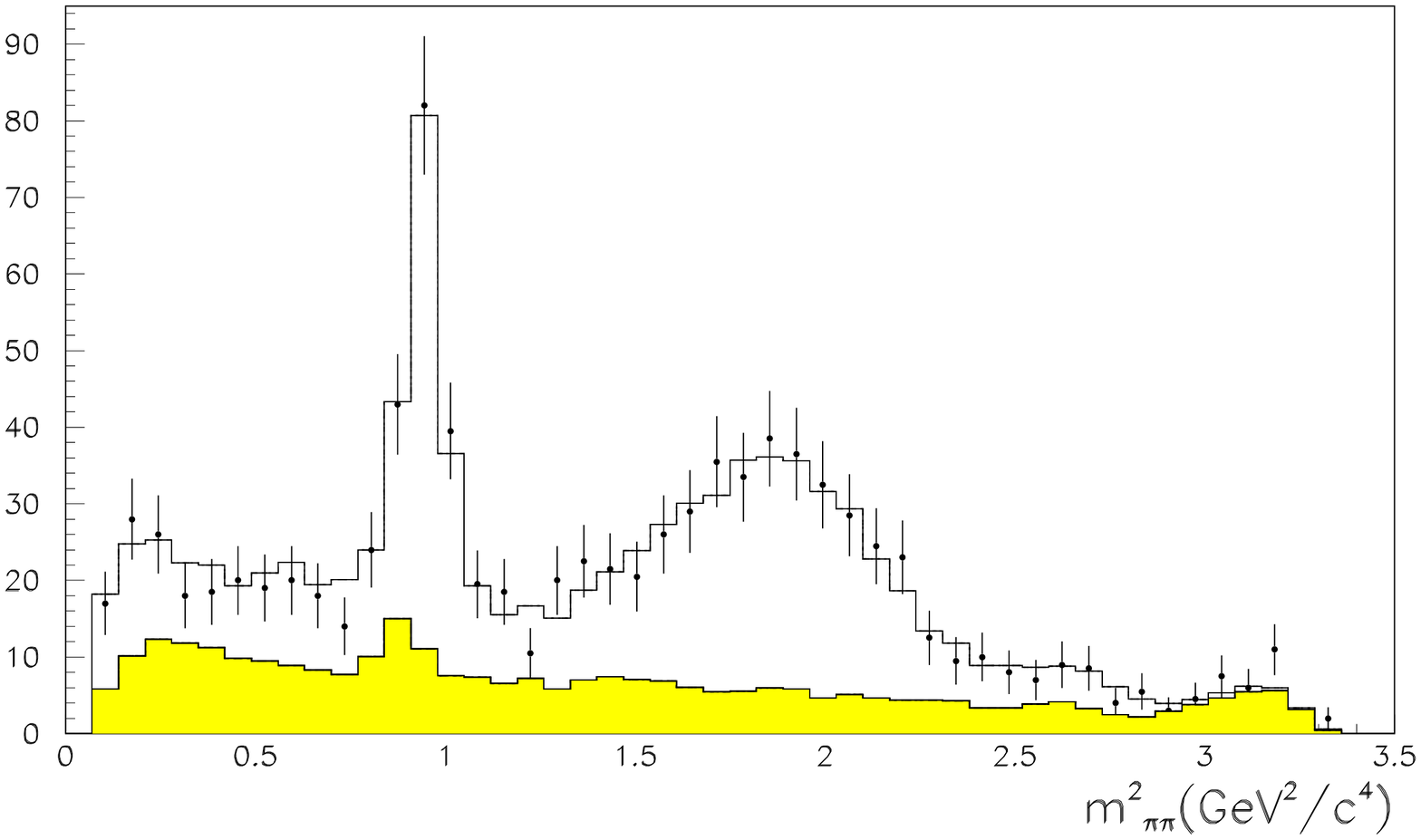,height=1.5in,width=2.in} \\
\end{tabular}
\end{center}
\vspace{10pt}
\caption{The $D_s^+ \rightarrow \pi^+ \pi^- \pi^+$ Dalitz plot and its 
projections ( preliminary results) from E791.}
\label{proyections2}
\end{figure}

The Dalitz plot of $D_s^+ \rightarrow \pi^- \pi^+ \pi^+$ and the
$\pi\pi$ mass projections \cite{Reis} are shown in Fig. \ref{proyections2}.

The $f_0 (980)\pi^{\pm}$ mode is the dominant one, accounting for nearly
half of the $D_s^+ \rightarrow \pi^- \pi^+ \pi^+$ decay width. The $f_0
(980)\pi^{\pm}$ is often supposed to have a large $s\bar s$ component,
indicating a large spectator amplitude in  
this decay. Significant contributions of $f_0 (1370)\pi^+$ and
$f_2(1270)\pi^+$ components were also found. The contribution of
$\rho^0(770)\pi^+$ and $\rho^0(1450)\pi^+$ components correspond to
about 10 $\%$ of the $\pi^- \pi^+ \pi^+$ width. This could indicate either
contribution from the annihilation diagram or from inelastic final
state interactions. No significant non-resonant component was found. \\
Preliminary $f_0$ masses and widths \cite{Reis} from E791 and PDG are presented
in the table \ref{masf0}.

\begin{table}
\caption{Preliminary results of $f_0$ mass and width of
$D_s^+ \rightarrow f_0 (980)\pi^+$ and $D_s^+ \rightarrow f_0 (1370)\pi^+$ 
from E791.}
\label{masf0}
\begin{tabular}{ccccc}
 &$f_0(980)$ & & $f_0(1370)$\\ 
& Mass (MeV/c$^2$)& width (MeV/c$^2$)& Mass (MeV/c$^2$)& width(MeV/c$^2$)\\
\tableline
E791 & $978 \pm 4$  & $44 \pm 5$  &  $1440 \pm 19$ & $165 \pm 29$ \\
PDG  & $978 \pm 10$ & $40 - 100$ & $1200 - 1500$ & $200 - 500$ \\
\end{tabular}
\end{table}

FOCUS also has preliminary results of this decay mode. Their
preliminary Dalitz plot  shows the
$D_s^+ \rightarrow \pi^- \pi^+ \pi^+$ based on a very clean signal of
$\sim$ 1300 events reconstructed from 100 $\%$ of their  data. The
preliminary results indicate a negligible contribution from the $\rho$
suggesting negligible Weak Annihilation contribution \cite{moroni}.

\subsubsection*{$D^+$ Dalitz plot results and evidence for a light scalar 
resonance}
\begin{figure}
\begin{center}
\begin{tabular}{ccc}
\epsfig{file=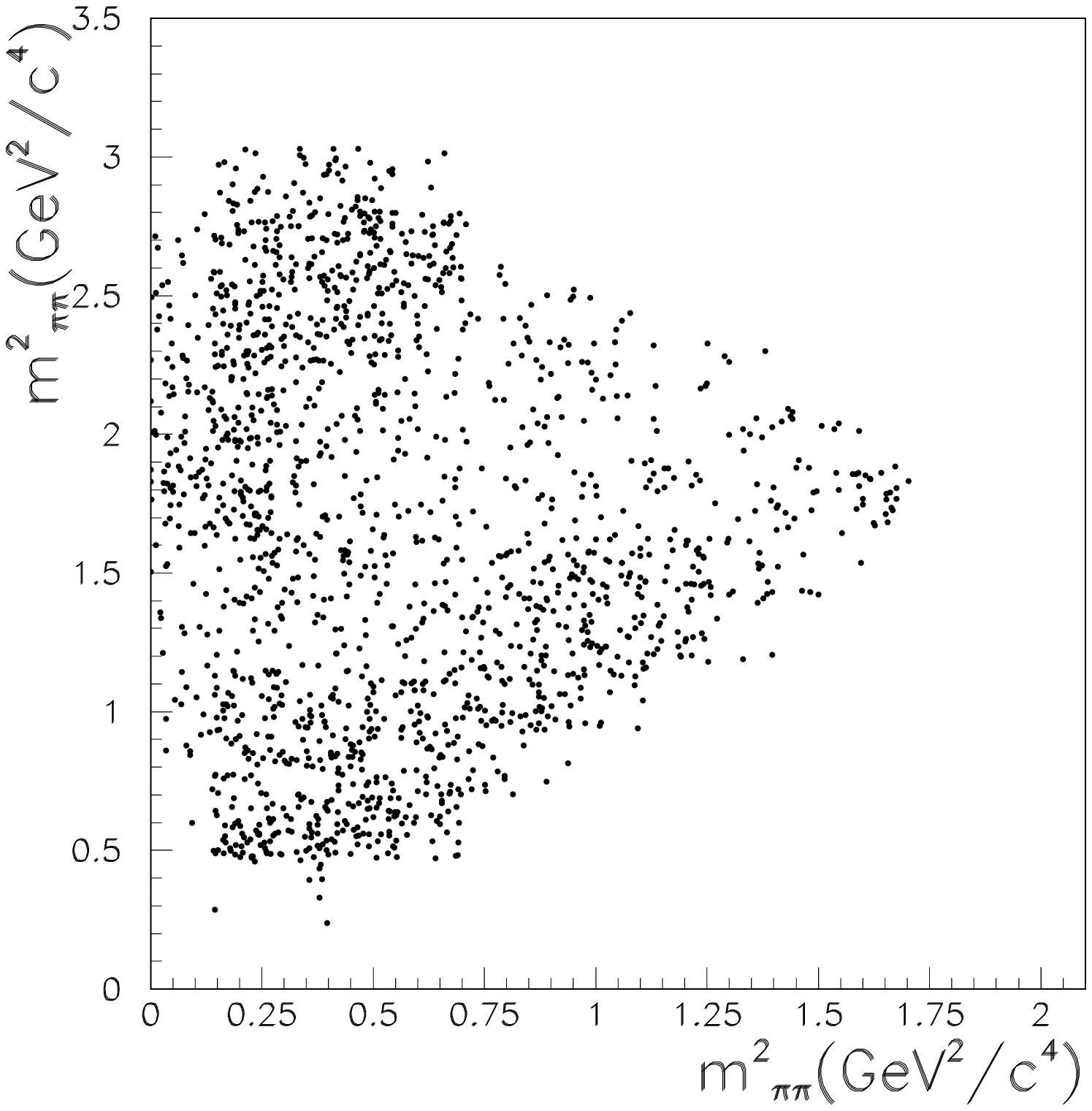,height=1.5in,width=1.5in}&  
\epsfig{file=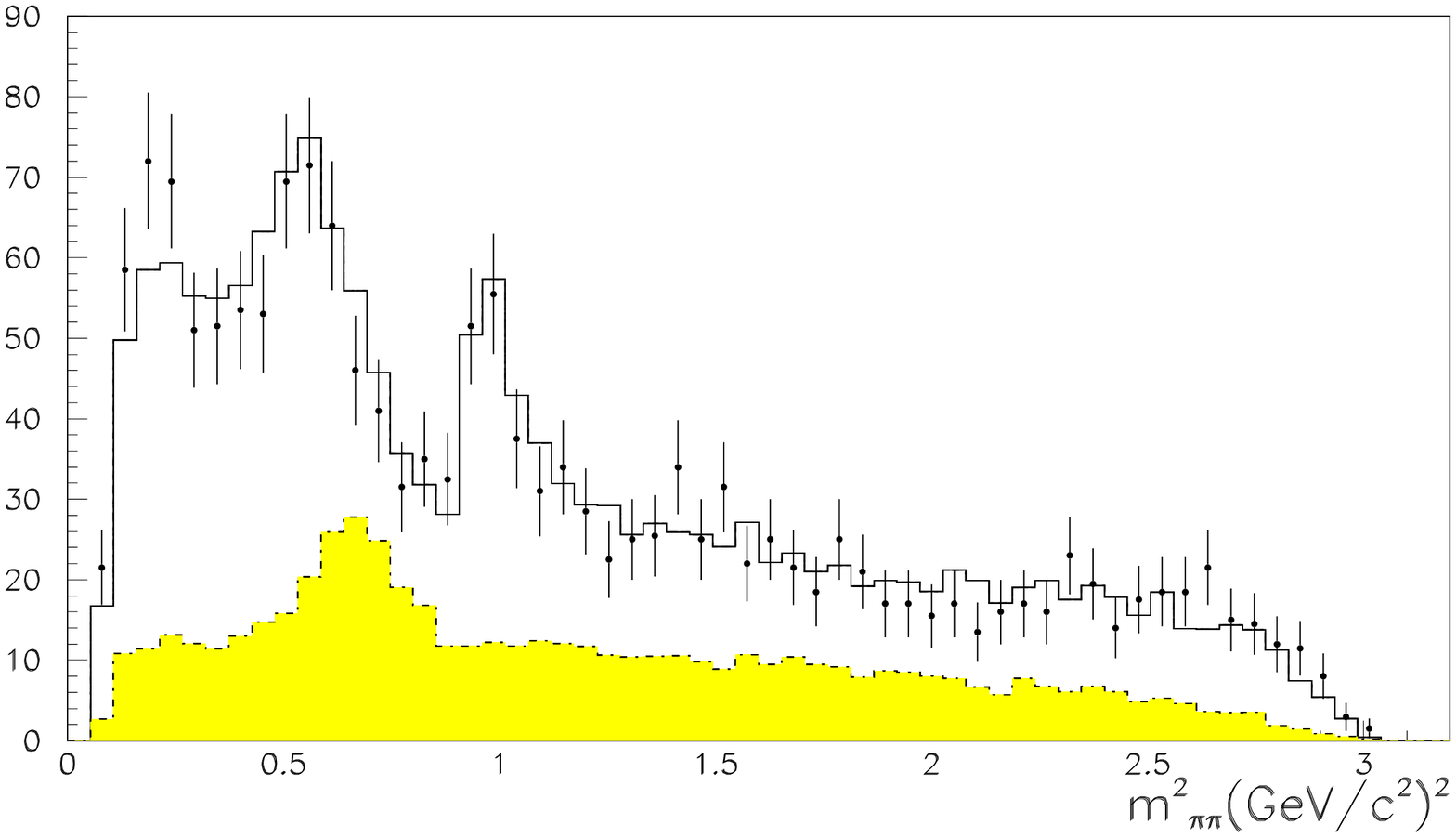,height=1.6in,width=1.7in}& 
\epsfig{file=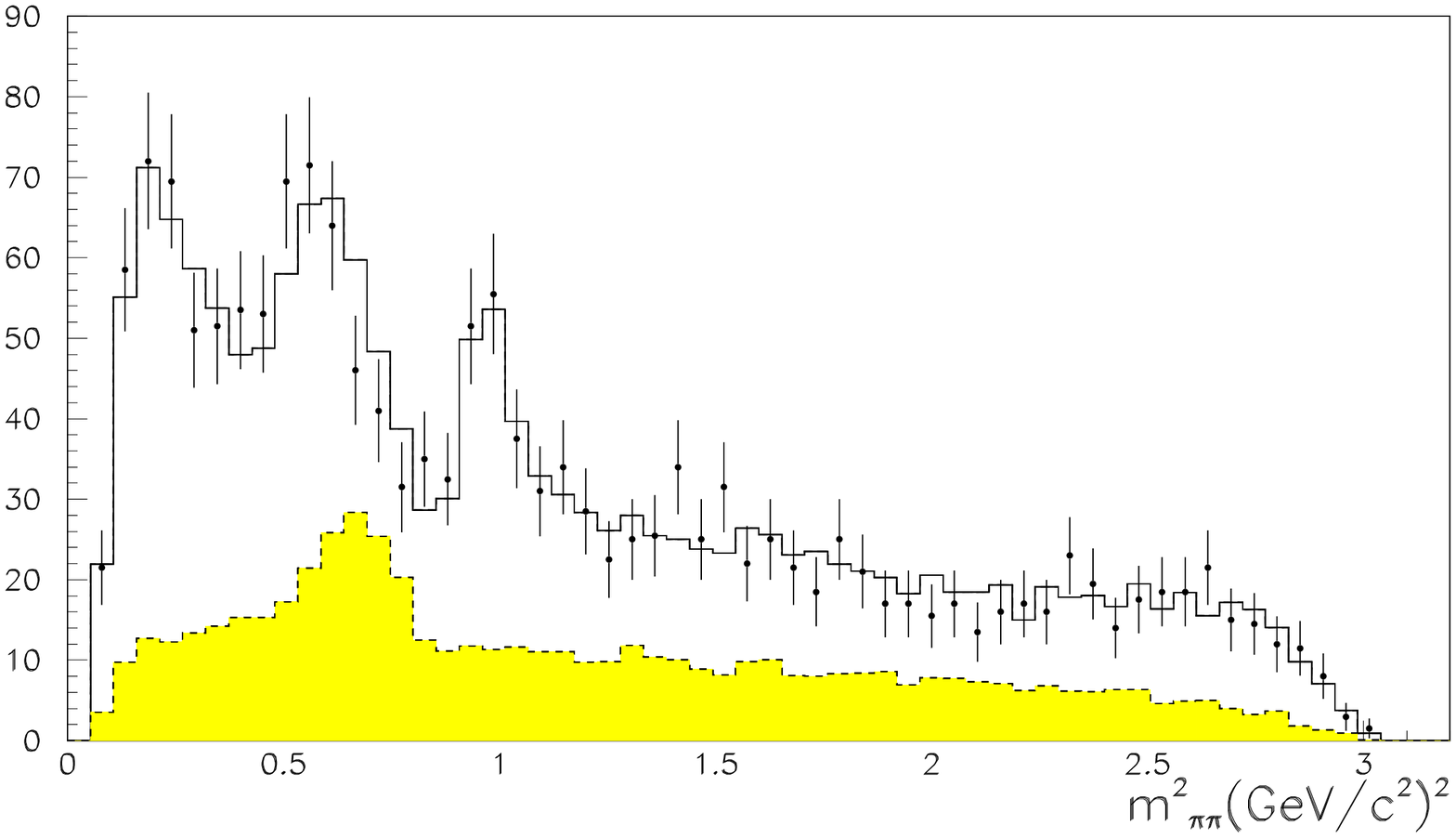,height=1.6in,width=1.7in} \\
\end{tabular}
\end{center}
\vspace{10pt}
\caption{$D^+ \pi^+ \pi^- \pi^+$ Dalitz plot  and its projections. 
The central plot is without a $\sigma \pi$ state and the last plot include a 
$\sigma \pi$ state in the fit (preliminary results). The shaded area is the 
background distribution.}
\label{proyections1}
\end{figure}

The Dalitz plot of the single Cabibbo-suppressed decay $D^+ \rightarrow
\pi^- \pi^+ \pi^+$ from E791 data is shown in Fig.\ref{proyections1} (left). A 
coherent amplitude analysis was used to determine the structure of its
density distribution. The fit including a non resonant amplitude and
amplitudes for $D^+$ decaying to a $\pi^+$ and any of the five
established $\pi^+ \pi^-$ resonances $\rho^0 (770)$, $f_0(980)$,
$f_2(1270)$, $f_0(1370)$, and $\rho^0(1450)$ is shown in Fig. 
\ref{proyections1} (central). This
fit is poor in the low $\pi^+ \pi^-$ region and has several
unsatisfactory features \cite{jussara}: the NR channel dominates, different 
from the $D_s^+$ decay, and the $\rho^0(1450) \pi^+$ is more significant than
the $\rho^0(770)\pi^+$ state. 

It was found that allowing an additional scalar state, with mass and
width unconstrained improves the fit substantially. The mass of the
resonance found by this fit is $486^{+28}_{-26}$ MeV/c$^2$ and the width
$351^{+51}_{-43}$ MeV/c$^2$. Referring to this $\pi^+ \pi^-$ resonance as the
$\sigma (500)$, it was found that $D^+ \rightarrow \sigma (500)\pi^+$
accounts for about half of the total decay rate, non-resonant decay was
very small and the $\rho^0 (1450) \pi^+$ fraction was much less than
$\rho^0 (770) \pi^+$. Preliminary results of the fit with this state
are shown in Fig. \ref{proyections1} (right side).

Theoretically, light scalar and isoscalar resonances are predicted in
models for spontaneous breaking of chiral symmetry, like the $\sigma$
linear model \cite{sigma-model}. These scalar particles have important
consequences for the quark model, for understanding low energy $\pi
\pi$ interactions and also for understanding the $\Delta I= 1/2$ rule.\\

\subsubsection*{Multidimensional analysis of 
$\Lambda_c^+ \rightarrow p K^- \pi^+$ from E791}

\begin{figure}
\begin{center}
\begin{tabular}{cccc}
\epsfig{file=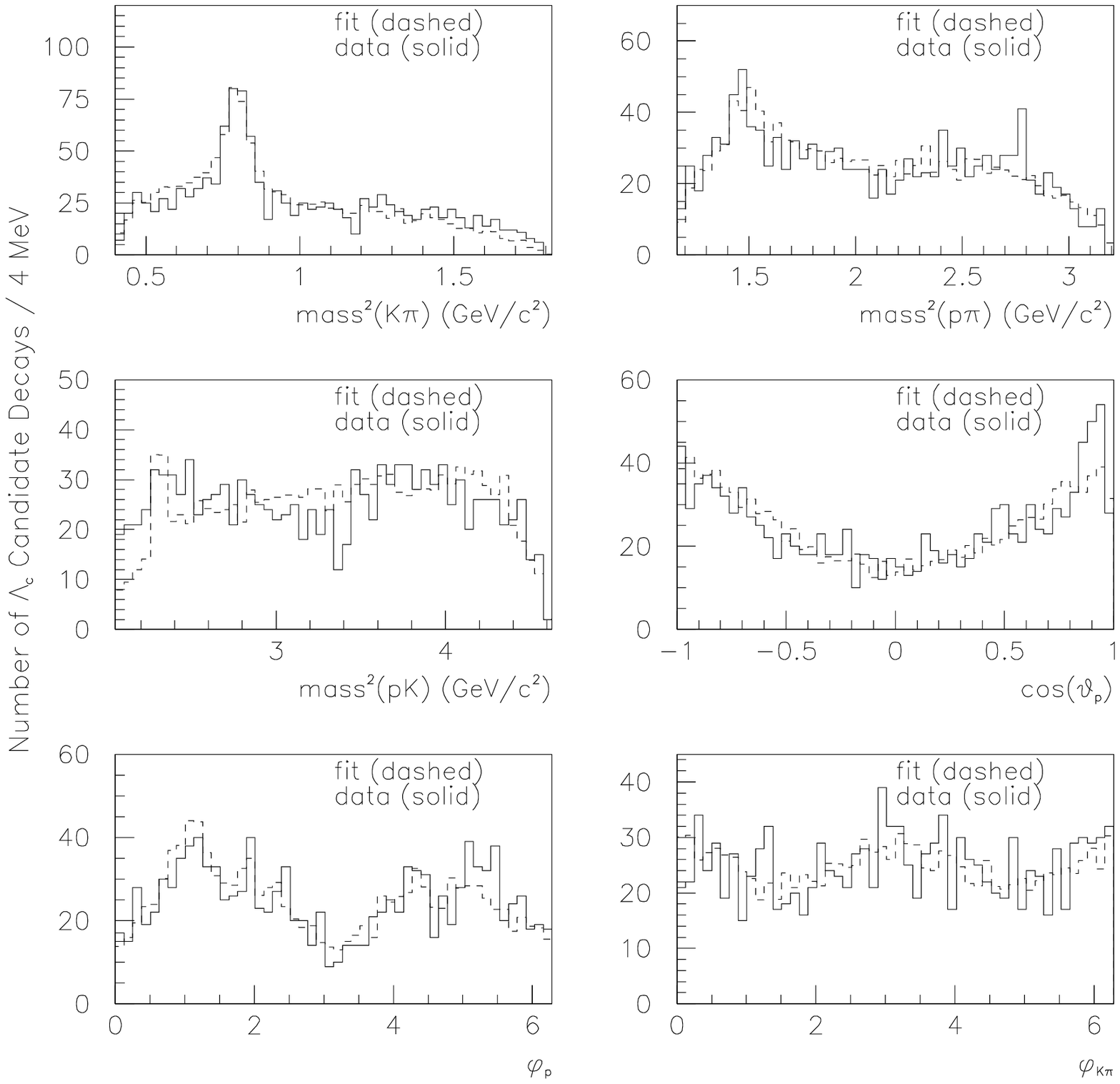,height=3in,width=3in} & & &
\epsfig{file=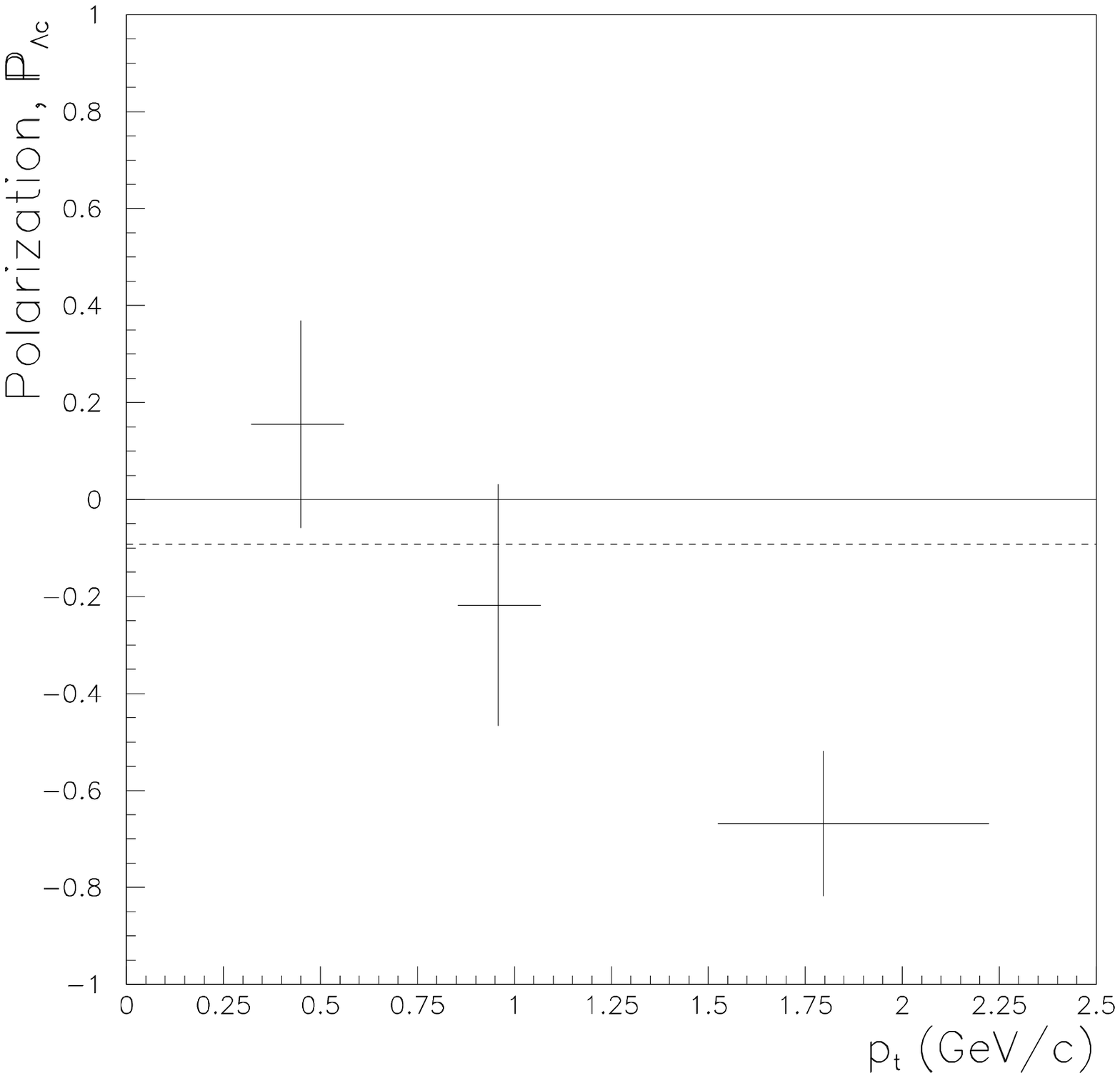,height=3in,width=1.7in} \\
\end{tabular}
\end{center}
\vspace{10pt}
\caption{ $\Lambda_c^+$ multidimensional analysis from E791: left plots shows 
projections of two body masses and angles. Solid 
line histograms are data in the range $2265 \:< \: M_{p K^- \pi^+} \: < \: 
2315 $ MeV/c$^2$. Dashed lines represent the fit in the same range. The right 
figure shows polarization $P_{\Lambda_c^+}$ vs. $p_t$.}
\label{resonLc}
\end{figure}

E791 has reported recently the first amplitude analysis of the decay of a
charm baryon \cite{e791-multia,brian-dcsd}. The study of charm baryon decays 
can give information 
regarding the relative importance of spectator and exchange amplitudes. 
Exchange amplitudes are small in charm meson decays because of helicity 
suppression. However in charm baryon decays this effect should not inhibit 
exchange amplitudes due to be three body nature of the interaction. The 
spectator and W-exchange diagrams can contribute to $pK^{*0}(890)$, 
$\Lambda (1520) \pi^+$ or $p K^- \pi^+$ modes. However for the $\Delta^{++} 
(1232)K^-$ the $W$-exchange is the only diagram possible.\\
The charm baryon can be produced polarized and its decay products carry 
spin. These extra quantum numbers require five kinematic variables for a 
complete description of the decay, and instead of the conventional two 
dimensional Dalitz plot analysis, a five-dimensional amplitude analysis 
is required.\\
A sample of $946 \pm 38$  $\Lambda_c^+ \rightarrow p K^- \pi^+$ 
reconstructed decays was used by E791 to determine relative strengths and
phases of resonances in the final state as well as the $\Lambda_c$ 
production polarization. The fit projections and the polarization as 
function of $p_t$ are shown in Fig.\ref{resonLc}. The resonant fractions 
for $\Lambda_c^+ \rightarrow p K^- \pi^+$ are shown in table \ref{reson}.\\
The $\Delta^{++} (1232)K^-$ and the $\Lambda (1520) \pi^+$ decay modes are 
seen as statistically significant for the first time. The observation of 
a substantial $\Delta^{++} (1232)K^-$ component provides strong evidence for 
the W-exchange amplitude in charm baryon decays. It was also observed an 
increasingly negative polarization for the $\Lambda_c$ as a function of 
$p_t$.

\begin{table}[!ht]
\caption{Resonant fractions for $\Lambda_c^+$ decay modes.}
\label{reson}
\begin{tabular}{lcc}
Decay Mode & \multicolumn{1}{c}{$\%$  of $ p K^- \pi^+$}\\
\tableline
$p K^{*0}(890)$           &  $ 19.5 \pm 2.6 \pm 1.8 $  \\
$\Delta^{++} (1232)K^-$   &  $ 18.0 \pm 2.9 \pm 2.9 $  \\
$\Lambda(1520) \pi^+$     &  $ 7.7 \pm 1.8 \pm 1.1 $   \\
non-resonant              &  $ 54.8 \pm 5.5 \pm 3.5 $  \\
\end{tabular}
\end{table}

\section*{Rare and forbidden decays}

\begin{figure}[!ht]
\centerline{\epsfig{file=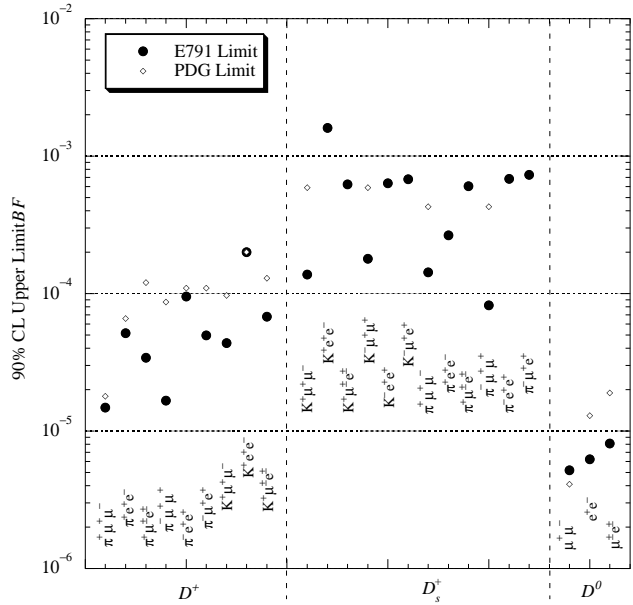,height=4.in,width=4.in}}
\centerline{\epsfig{file=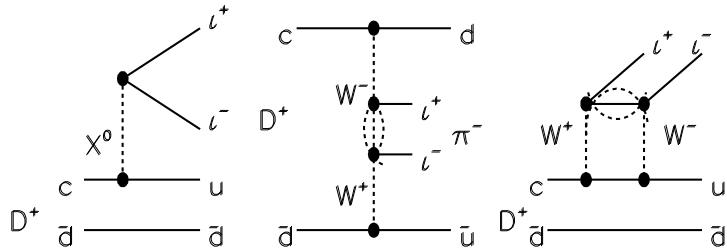,height=1.5in,width=4.in}}
\vspace{10pt}
\caption{Rare decays limits from E791 (top) and some Feynman diagrams 
for FCNC, LFNV and LNV decay modes (bottom).}
\label{rdecays}
\end{figure}

In the charm sector the rare and forbidden dilepton decay modes can be 
classified mainly into three categories (example of Feynman diagrams are 
shown in Fig. \ref{rdecays}):

\noindent
1) Flavor Changing Neutral Current decays (FCNC) such as 
$D^0 \rightarrow l^+ l^-$ and $D^+ \rightarrow h^+ l^+ l^-$.\\
\noindent
2) Lepton Family Number Violating decays (LFNV) such as 
$D^+ \rightarrow h^+ l_1^+ l_2^-$ and $D^0 \rightarrow l_1^+ l_2^-$ where 
the leptons are from different generations.\\ 
\noindent
3) Lepton number Violating decays (LNV) such as $D^+ \rightarrow h^- l^+ 
l^+$  where the leptons are of the same generation but have the same 
sign.\\
\noindent
Where $h$ stands for $\pi$, $K$ and $l$ for $e, \mu$.

\noindent
The first decay modes (FCNC) are rare, that means a process suppressed 
via the GIM mechanism which proceeds via an internal quarks loop in the 
Standard Model \cite{scwartz}. 
The FCNC decay mode $D^0 \rightarrow l^+ l^-$ can proceed via a $W$ box 
diagram and the theoretical estimates \cite{Hewett} for the branching 
fraction are of the order of $\sim 10^{-19}$. The predictions for the 
other 
FCNC decay modes, $D^+ \rightarrow h^+ l^+ l^-$, are considerably larger, 
of the order of $\sim 10^{-9}$ . These decay modes can proceed via 
penguin diagrams \cite{scwartz} and from long  distance effects.\\
The decays modes LFNV and LNV are strictly forbidden in the Standard 
Model as they do not conserve lepton number. However, some theoretical 
extensions of the Standard Model predict lepton number violation 
\cite{NVL}, and then the observation of a signal in these modes would be 
evidence for new physics beyond the SM.\\

E791 has recently published \cite{e791-rares} (see figure \ref{rdecays}) a 
set of new limits in rare 
or forbidden decay modes that improve  the PDG98 numbers by a factor of 
10. They searched for 24 different rare and forbidden decay modes and 
have found no evidence for them. They therefore presented upper limits on 
their branching fractions. Fourteen of their limits represent a 
significant improvement over previous results and eight are presented for 
the first time.

For this study E791 used a {\it blind } analysis technique. The mass region 
where the signal is expected is {\it masked} throughout the analysis. 
Selection criteria are optimised by studying signal events generated by 
Monte Carlo simulation and background events obtained from data in mass 
windows above and below the {\it signal region}. The criteria were chosen to 
maximize the ratio $N_S / \sqrt(N_B)$, where $N_S$ and $N_B$ are the 
numbers of signal and background events, respectively. Only after this 
procedure were events within the signal window unmasked. This blind 
technique is used so that the presence or absence of signal does not bias 
the choice of the selection criteria. 

FOCUS is also looking for rare and forbidden decays using the same 
technique. Some examples of rare decays are presented in table 
\ref{factor1} where we compare the expected sensitivity from FOCUS to E791 
results and PDG values. 

\begin{table}[!ht]
\caption{ Rare decays limits from FOCUS (preliminary), E791 and PDG98.}
\label{factor1}
\begin{tabular}{lddd}
   Decay mode& FOCUS expected sensitivity &
   \multicolumn{1}{c}{E791\tablenote{Published \cite{e791-rares}}} 
&
  \multicolumn{1}{c}{PDG 98}\\
 &(45 $\%$ of the data) & 90 $\%$ C.L. limit & 90 $\%$ C.L limit \\
\tableline
$D^+ \rightarrow K^+ \mu^+ \mu^-$ & 8.1 $\times 10^{-6}$ 
& 4.4 $\times 10^{-5}$ & 9.7 $\times 10^{-5}$. \\ 
$D^+\rightarrow K^-\mu^+ \mu^+$ & 12.1 $\times 10^{-6}$ 
&   &  1.2 $\times 10^{-4}$. \\ 
$D^+\rightarrow \pi^+\mu^+ \mu^-$ & 7.8 $\times 10^{-6}$ 
& 1.5 $\times 10^{-5}$& 1.8 $\times 10^{-5}$. \\
$D^+\rightarrow \pi^-\mu^+ \mu^+$ & 7.1 $\times 10^{-6}$ 
& 1.7 $\times 10^{-5}$ & 8.7 $\times 10^{-5}$. \\ 
$D^+\rightarrow \mu^- \mu^+ \mu^+$ & 4.4 $\times 10^{-6}$ 
&  &  \\
\end{tabular}
\end{table}

\section*{Semileptonic decays, Form-factors} 

The weak decays of hadrons containing heavy quarks are
influenced by strong interaction effects. Semileptonic charm decays such
as $D^+ \rightarrow \overline {K}^{*0} e^+ \nu_e$, 
$D_s^+ \rightarrow \phi l^+ \nu_l$ are an
especially clean way to study these effects because the leptonic and
hadronic currents completely factorize in the decay amplitude ,$A$, as we can 
see  in the Eq. \ref{amplitude},  where $G_F$ is the Fermi coupling constant 
for the weak interaction and $V_{cs}$ is the CKM matrix element. $L^{\mu}$  
(Eq. \ref{leptonic})  and  $H_{\mu}$ (Eq. \ref{hadronic}) represent the 
leptonic and hadronic currents \cite{key3}.\\

\begin{equation}
A(D^+ \rightarrow \overline {K^*}^0 e^+ \nu_e)= \frac{G_F}{\sqrt{2}} V_{cs}
L^{\mu} H_{\mu}
\label{amplitude}
\end{equation}

\begin{equation}
L^{\mu} = \bar u_e \gamma^{\mu} (1-\gamma_5)v_{\nu}
\label{leptonic}
\end{equation}

\begin{eqnarray}
H_{\mu}=(m_D+m_{K^*}) A_1(q^2)\epsilon_{mu}
-\frac{A_2(q^2)}{m_D+m_{K^*}}(\epsilon \cdot p_D)(p_D+p_K)_{\mu}-
\nonumber \\
\frac{A_3(q^2)}{m_D+m_{K^*}}(\epsilon \cdot p_D)(p_D-p_K)_{\mu}-
i \frac{2 V(q^2)}{m_D+m_{K^*}}\varepsilon_{\mu \nu \rho \sigma}
\epsilon^{\nu}p_D^{\rho}p_K^{\sigma} 
\label{hadronic}
\end{eqnarray}

With a vector meson in the final state, there are four form factors,
$V(q^2)$, $A_1(q^2)$, $A_2(q^2)$ and $A_3(q^2)$, which are functions of
the Lorentz-invariant momentum transfer squared $q^2$, the square of the 
invariant mass of the virtual W  \cite{key3}.  
The differential decay rate for $D^+ \rightarrow \overline{K}^{\,\star 0}
\mu^+ \nu_\mu$ with $\overline{K}^{\,\star 0} \rightarrow K^- \pi^+$ is a
quadratic homogeneous function of the four form factors. Unfortunately,
the limited size of current data samples does not allow precise 
measurement
of the $q^2$-dependence of the form factors; we thus assume the 
dependence to
be given by the nearest-pole dominance model:
$F(q^2) =F(0)/(1-q^2/m_{pole}^2)$ 
where $m_{pole} = m_V = 2.1 \: {\rm GeV}/c^2$ for the vector form factor 
$V$ (which correspond to $J^P\:=\:1^+$ state, $D^*_{s1}$), and 
$m_{pole} = m_A = 2.5 \: {\rm GeV}/c^2$ for the three axial-
vector form factors A \cite{key4} (corresponding to $1^{-}$ state, $D^*_s$).\\ 
The third form factor $A_3(q^2)$, which is unobservable in the limit of 
vanishing lepton mass, probes the spin-0 component of the off-shell $W$. 
Additional spin-flip amplitudes, suppressed by an overall factor of 
$m_{\ell}^2/q^2$ when compared with spin no-flip amplitudes, contribute 
to the differential decay rate.  
Because $A_1(q^2)$ appears among the coefficients of every
term in the differential decay rate, we can factor out
$A_1(0)$ and measure the ratios: \\ 
$r_V = V(0)/A_1(0)$, $r_2 = A_2(0)/A_1(0)$ and $r_3 = A_3(0)/A_1(0)$. 
The values of these ratios can be extracted without any assumption about 
the total decay rate or the weak mixing matrix element $V_{cs}$. \\

\begin{figure}[!ht]
\begin{center}
\begin{tabular}{cccc}
\epsfig{file=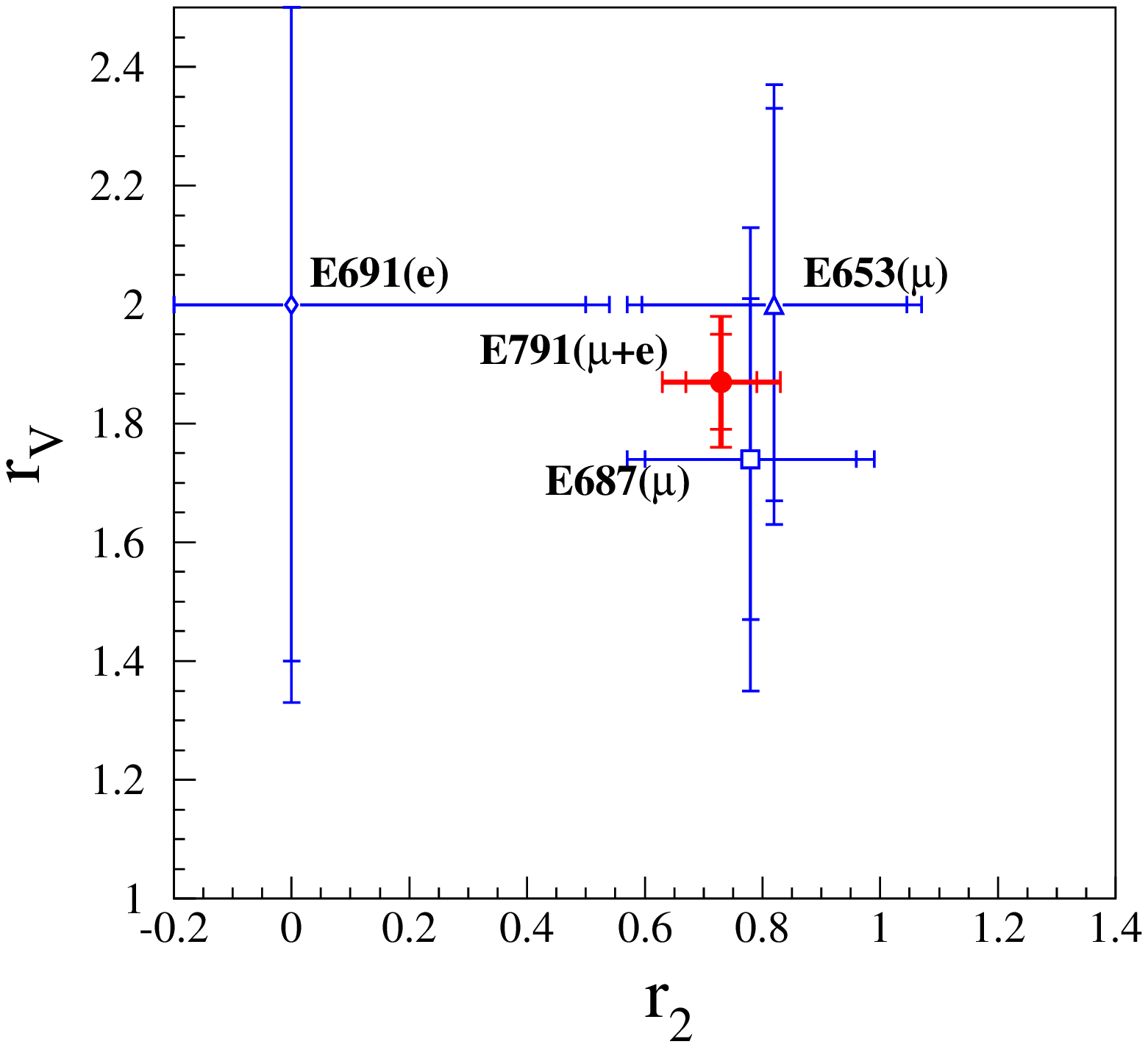,height=2in,width=2in}& &
\epsfig{file=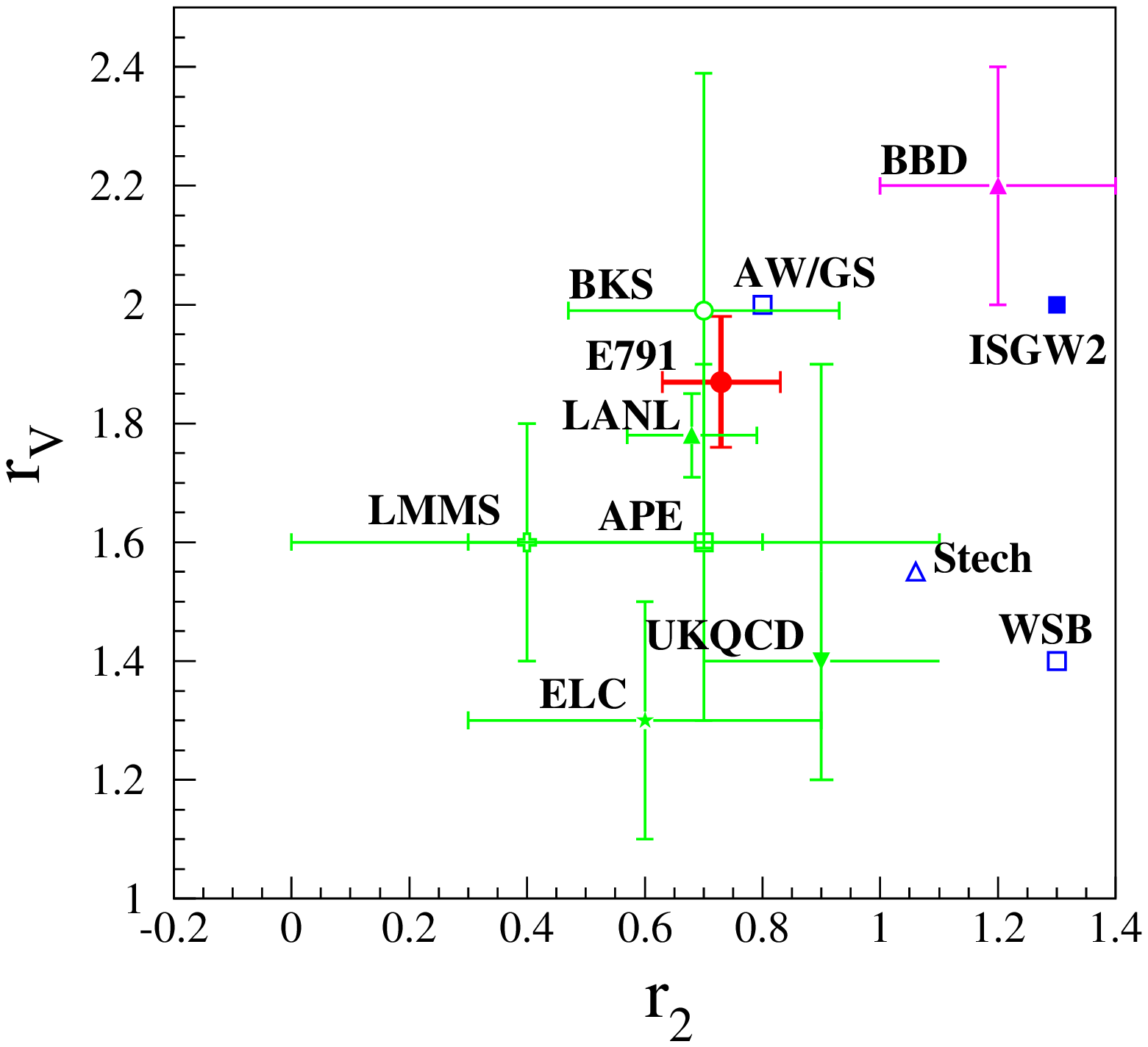,height=2in,width=2in}\\
\epsfig{file=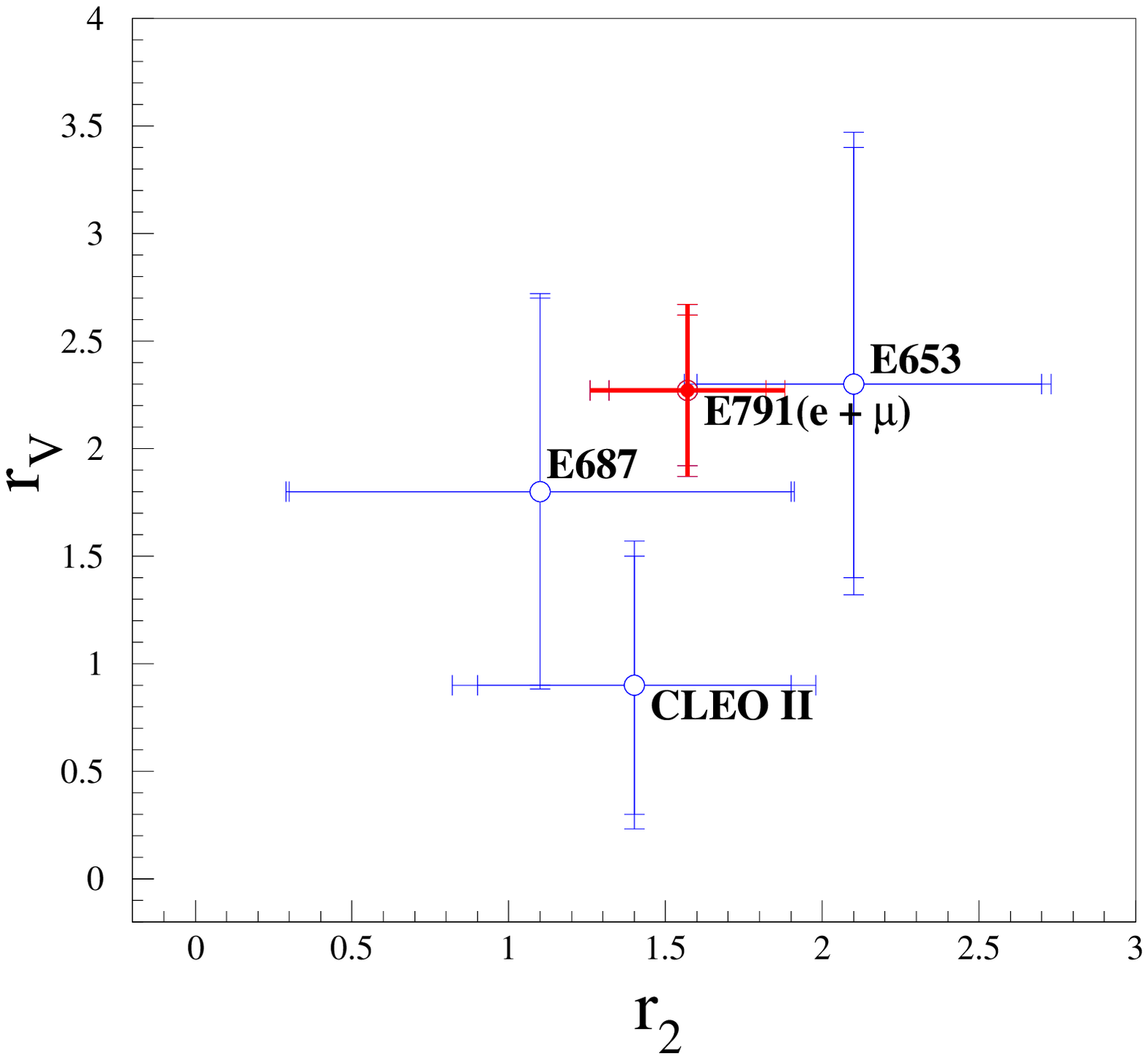,height=2in,width=2in}& &
\epsfig{file=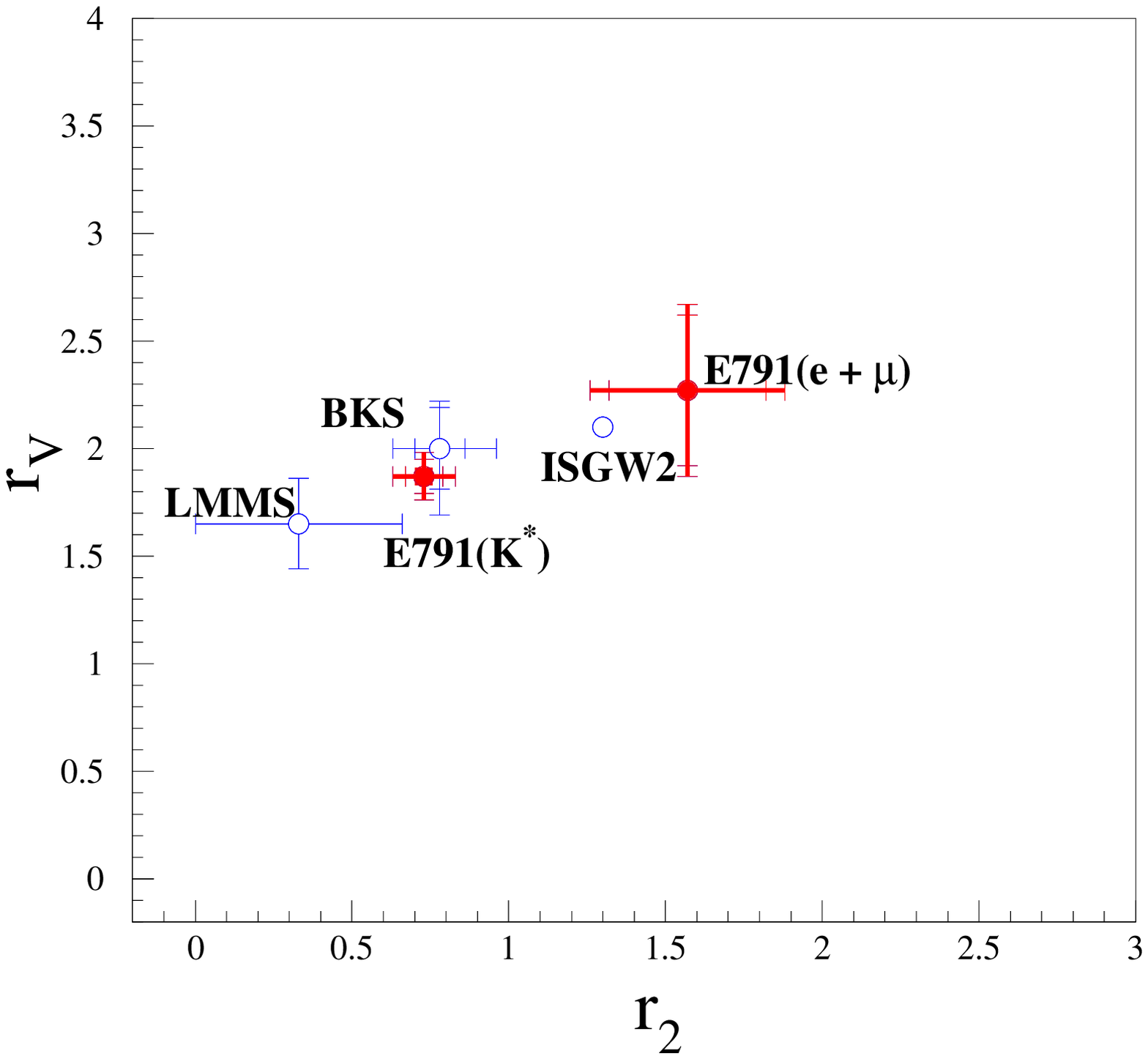,height=2in,width=2in} \\
\end{tabular}
\end{center}
\vspace{10pt}
\caption{Top-left figure present the ratios $r_2$ and $r_V$ for 
$D^+ \rightarrow \overline{K^{*0}} l \nu_l$, and 
top-right compare them to theoretical models. Bottom-left show ratios $r_2$ 
and $r_V$ for $D_s^+ \rightarrow \overline{\phi} l \nu_l$, and bottom-right 
is a comparison with theoretical models.}
\label{fig14}
\end{figure}

\noindent
We report E791 measurements of the form factor ratios $r_v$ and $r_2$ for the 
muon channel \cite{factor1-791} 
and combined with electron channel \cite{factor2-791} (see Fig. \ref{fig14}). 
This is the first set of measurements in both muon and electron channels 
from a
single experiment. We also report the first measurement of $r_3 =
A_3(0)/A_1(0)$, which is unobservable in the limit of vanishing charged
lepton mass. \\
The measurements of the form factor ratios for $D^+ \rightarrow
\overline{K}^{\,\star 0} \mu^+ \nu_\mu$ presented here and for the 
similar
decay channel $D^+ \rightarrow \overline{K}^{\,\star 0} e^+ \nu_e$
\cite{key5} follow the same analysis procedure except for the charged
lepton identification. Both results in the electron and muon channels are 
consistent within errors, supporting the assumption that strong interaction 
effects incorporated in the values of form factor ratios do not depend on 
the particular $W^+$ leptonic decay. \\
The combined results of electronic and muonic decay modes produce 
$r_V = 1.87 \pm 0.08 \pm 0.07$ and
$r_2 = 0.73 \pm 0.06 \pm 0.08$. The combination
of all systematic errors is ultimately close to that which one would
obtain assuming all the errors are uncorrelated. The third form factor
ratio $r_3 = 0.04 \pm 0.33 \pm 0.29$ was not measured in the electronic 
mode. \\
Table~\ref{formfactor} compares the values of the form factor ratios $r_V$ 
and $r_2$ measured by E791 in the electron, muon and combined modes with
previous experimental results. The size of the data sample and the decay
channel are listed for each case. All experimental results are consistent
within errors. The comparison between the E791 combined values of the 
form factor ratios $r_V$ and $r_2$ and other experimental results is also
shown in Fig. \ref{fig14} together with theoretical predictions.

\begin{table}[!ht]
\caption{Comparison of E791 results with previous  results.}
\label{formfactor}
\begin{center}
\begin{tabular}{llll}
Exp. & Events & $r_V = V(0)/A_1(0)$ & $r_2 = A_2(0)/A_1(0)$  \\ \hline
E791 & 6000 ($e+\mu$) & $1.87 \pm 0.08 \pm 0.07$ & $0.73 \pm 0.06 \pm 0.08$ \\
E791 &  3000 ($\mu$) & $1.84 \pm 0.11 \pm 0.09$ & $0.75 \pm 0.08 \pm 0.09$ \\
E791 & 3000 ($e$) & $1.90 \pm 0.11 \pm 0.09$ & $0.71 \pm 0.08 \pm 0.09$ \\
E687\cite{f1687} & 900 ($\mu$) & $1.74 \pm 0.27 \pm 0.28$ & $0.78 \pm 0.18 
\pm 0.10$ \\
E653\cite{f1653} & 300 ($\mu$) & $2.00^{+0.34}_{-0.32} \pm 0.16$ & 
$0.82^{+0.22}_{-0.23} \pm 0.11$ \\
E691\cite{f1691} & 200 ($e$) & $2.0 \pm 0.6 \pm 0.3$ & $0.0 \pm 0.5 \pm 0.2$ \\
\end{tabular}
\end{center}
\end{table}

The FOCUS experiments anticipate measuring $r_2$ and $r_V$ to better 
precision than previous experiments \cite{BrianSL}

 \section*{Charm Lifetimes}

The study of the charm particle lifetimes is motivated by two main goals: 
to extract partial decay rates and to study decay dynamics.
The total decay width can be expressed as a sum of the three possible 
classes of decays, so the lifetime of a particle can be written as

\begin{equation}
\tau ={{\hbar}\over{\Gamma_{tot}}} =
{{\hbar}\over{\Gamma_{leptonic}+\Gamma_{SL}
+\Gamma_{nonleptonic}}}
 \label{tau} 
\end{equation}

\begin{figure}[!ht]
\centerline{\epsfig{file=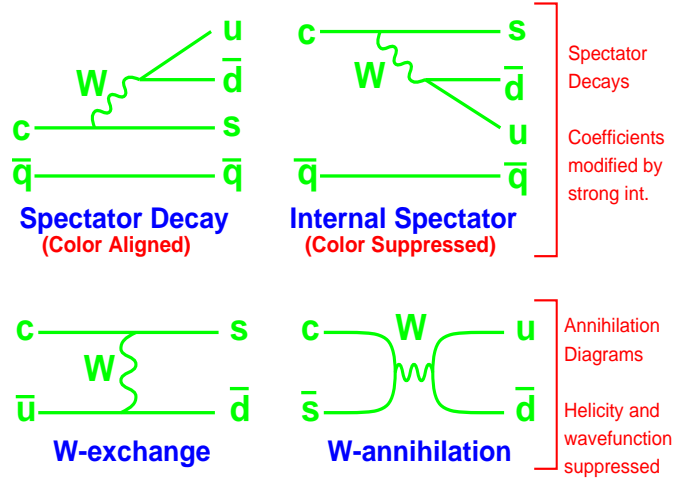,height=2.5in,width=3.5in}}
\vspace{10pt}
\caption{Hadronic decays diagrams for charm meson decays.}
\label{hadro-decays}
\end{figure}

\noindent
The leptonic partial width is normally very small ($\Gamma_{leptonic}  
\sim 10^{-3} \: - 10^{-4}$) due to helicity suppression. The observed 
$\Gamma_{SL}(D^+)$ and
$\Gamma_{SL}(D^0)$ are equal to within $10\%$, as expected from isospin 
invariance.  So, the large difference observed in the $D^+$ and $D^0$ 
lifetimes, $\tau (D^+)/\tau (D^0) = 2.55\pm 0.04$, is due to a large 
difference in the hadronic decay rates ($\Gamma_{nonleptonic}$) for the 
$D^+$ and the $D^0$. If there were no other diagram but the spectator and 
no QCD effects we would have the same value for $\tau$. Thus, the 
different lifetimes are an indication that we need to take into account 
contributions from diagrams where the W interact with two valence quarks, 
such as W-annihilation (WA) and W-exchange (WX), and any interference 
between them, as shows the Fig. \ref{hadro-decays}.\\ 

A systematic approach now exists for the treatment of inclusive decays, 
based on QCD and consists of an Operator Product Expansion (OPE) in the 
Heavy Quark mass. In this approach the interaction is factorized into 
three parts: weak interaction between quarks, perturbative QCD 
corrections and non-perturbative QCD effects. The decay rate is given by

\begin{equation}
\Gamma_{HQ}= \frac{G_F^2m_Q^2}{192 \pi^3} \Sigma f_i|V_{Qq_i}|^2 
[A_1 + \frac{A_2}{\Delta^2} + \frac{A_3}{\Delta^3}+ \ldots ]
\end{equation}

\noindent
where $\Delta$ is taken as the heavy quark mass and $f_i$ is a phase 
space factor. $A_1=1$ contains the spectator diagram contribution; $A_2$ 
is the spin interaction of the heavy quark with light quark degrees of 
freedom inside the hadron and produces differences between the baryon and 
meson lifetimes; $A_3$ 
includes the non-spectator $W$-annihilation, $W$-exchange and Pauli 
Interference 
(PI) of the decay and the spectator quarks contributions.\\

New results on charm lifetime measurements are shown in table \ref{taus}. 
The most relevant information is the $D_s^+$ 
lifetime from E791 \cite{tau-Ds,taud0-791} and FOCUS \cite{tau-Ds-focus}
which is now conclusively above the $D^0$ lifetime, with a ratio

\begin{equation}
R_\tau = \tau (D_s^+)/\tau (D^0) = 1.22\pm 0.02
\end{equation}
It is important to note that $R_{\tau}$ is now ten standard deviations 
away from unity, indicating that although not dominant, the WA/WX 
contributions are significant. In fact the OPE model predicts $R_{\tau}$ 
= 1.00 - 1.07 without WA/WX contributions, allowing a variation of $\pm 
20 \%$ if the WA/WX operators are included, in agreement with the new 
result.
 
\begin{figure}[!ht] 
\begin{center}
\begin{tabular}{cc}
\epsfig{file=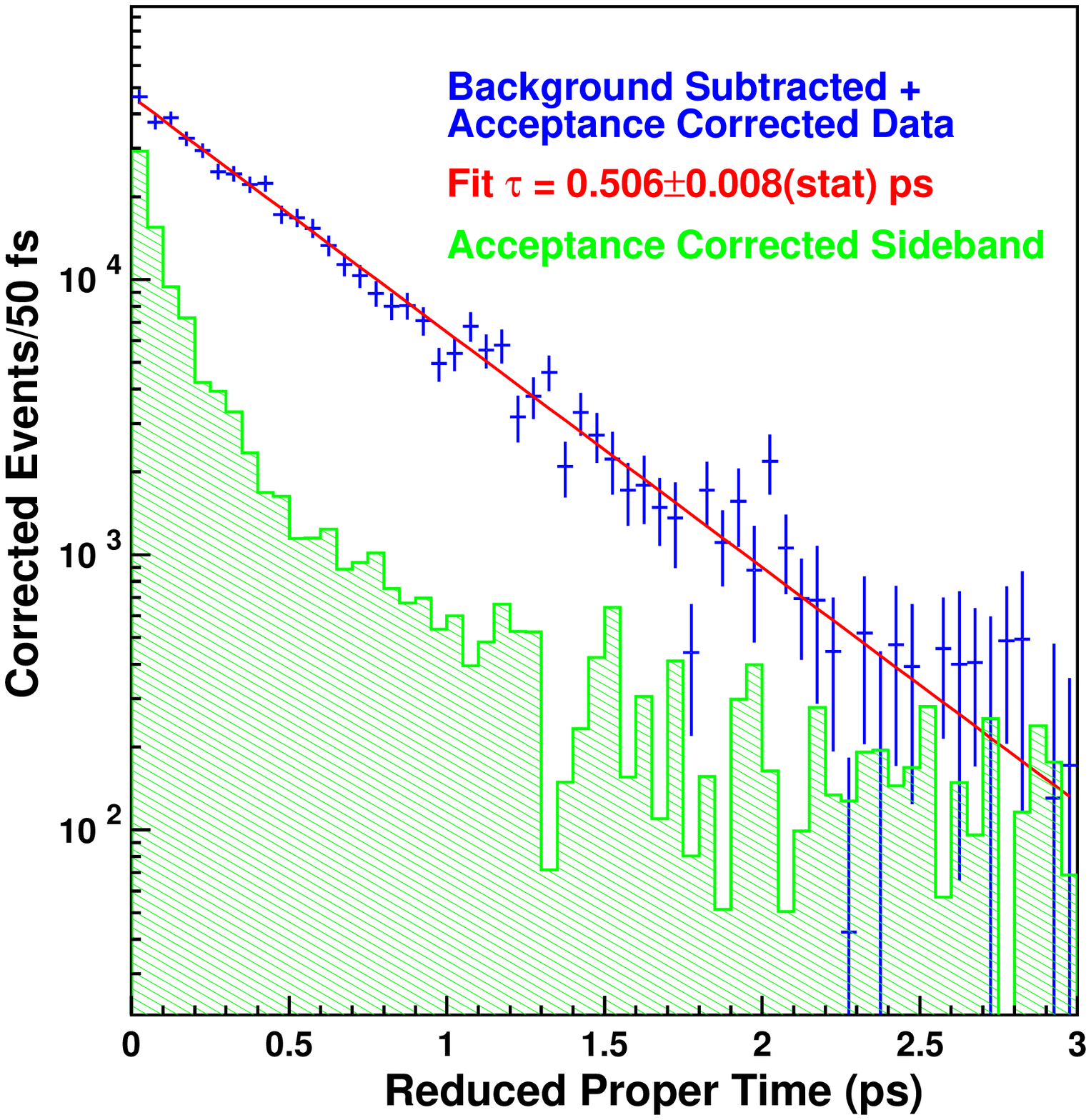,height=2.7in,width=2.6in}&
\epsfig{file=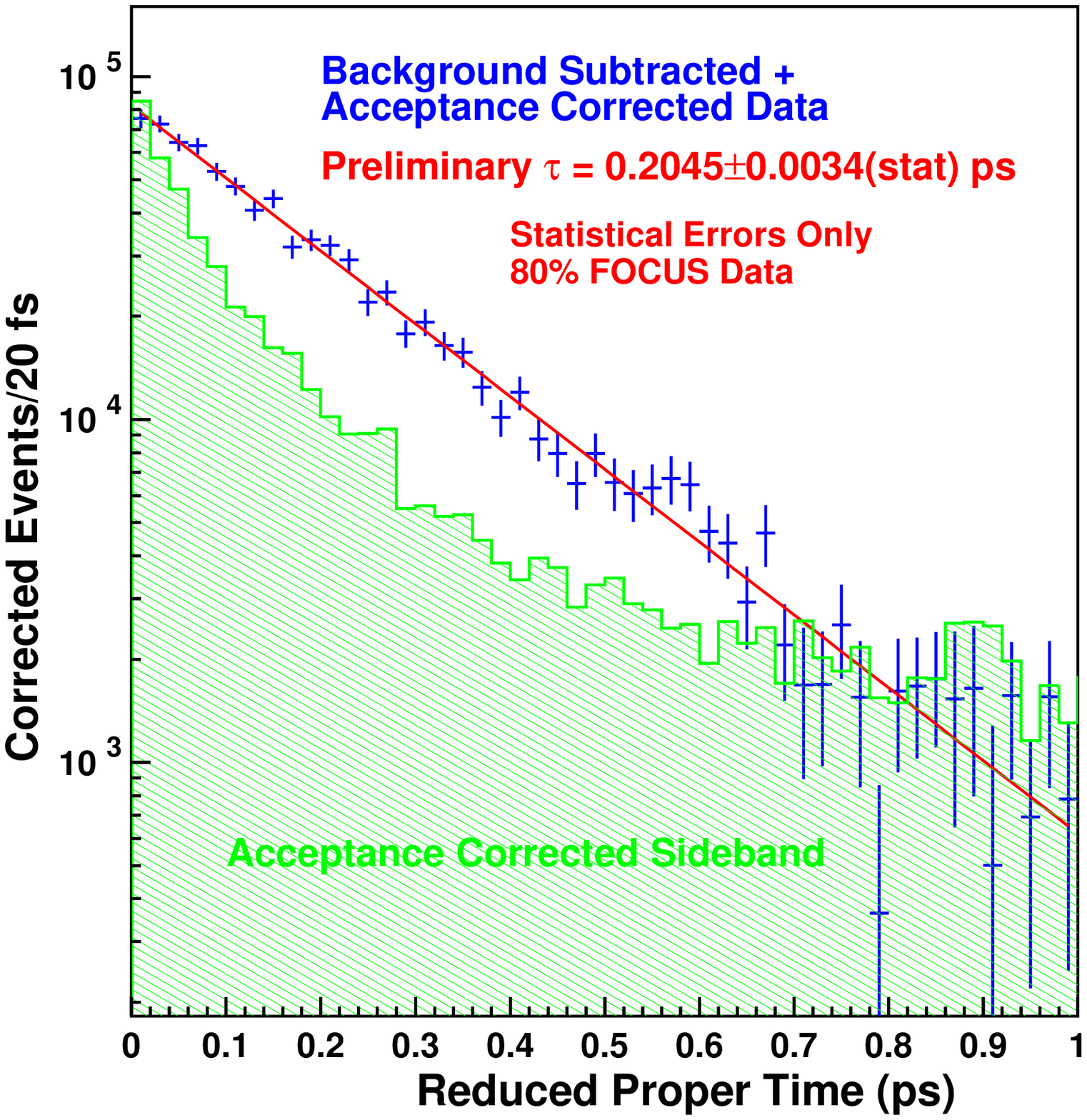,height=2.5in,width=2.5in}\\
\end{tabular}
\end{center}
\vspace{10pt}
\caption{$D_s^+$  (left) and $\Lambda_c^+$ (right) lifetimes from FOCUS 
(preliminary results)}
\label{fig17}
\end{figure}

In the baryon sector SELEX  and FOCUS have preliminary measurements of the 
$\Lambda_c^+$ lifetime, as shown in table.\ref{taus} and Fig. \ref{fig17}.  
Using 100$\%$ of their 
data  in $\Lambda_c^+ \rightarrow p K^- \pi^+$, SELEX find $\tau =177\pm 10 
(stat) fs$ \cite{tau-lambda-selex}, in a $2-3\sigma$ disagreement with PDG98 
and FOCUS.

\begin{table}[!ht] 
\caption{Summary of new charm lifetime measurements.}
\label{taus}
\begin{tabular}{lcccccc}
Experiment & $\tau (D^+)$ fs & $\tau (D^0)$ fs & $\tau (D_s^+)$ fs&  
$\tau (\Lambda_c^+)$ fs\\
\tableline
PDG'98 & $1057 \pm 15$  & $415 \pm 4$ & $467 \pm 17$ & $206 \pm 12$   \\
E791\tablenote{$\tau (D^+)$ using only the $\phi \pi^+$ mode.} 
       & $1065 \pm 48$ & $ 413 \pm 3 \pm 4$ & $ 518 \pm 14 \pm 7$ &    \\ 
CLEO   & $1033.6 \pm 22.1^{+9.9}_{-12.7}$ & $408 \pm 4.1^{+3.5}_{-3.4}$ &
       $486.3 \pm 15.0^{+4.9}_{-5.1}$ & \\
FOCUS$^b$  & & & $ 506 \pm 8$ & $ 204.5 \pm 3.4 $ \\
SELEX\tablenote{Preliminary results with no systematic uncertainty quoted}
       & & & & $177 \pm 10$\\ 
\tableline \tableline
World Average & $1052 \pm 12 $ & $412.8 \pm 2.7$ & $499.9 \pm 6.1$ & 
$201.9 \pm 3.1$ \\
\end{tabular}
\end{table}

\subsection*{Lifetime differences and $D^0 \overline {D^0}$ mixing}

\begin{figure}[!ht] 
\begin{center}
\begin{tabular}{cc}
\epsfig{file=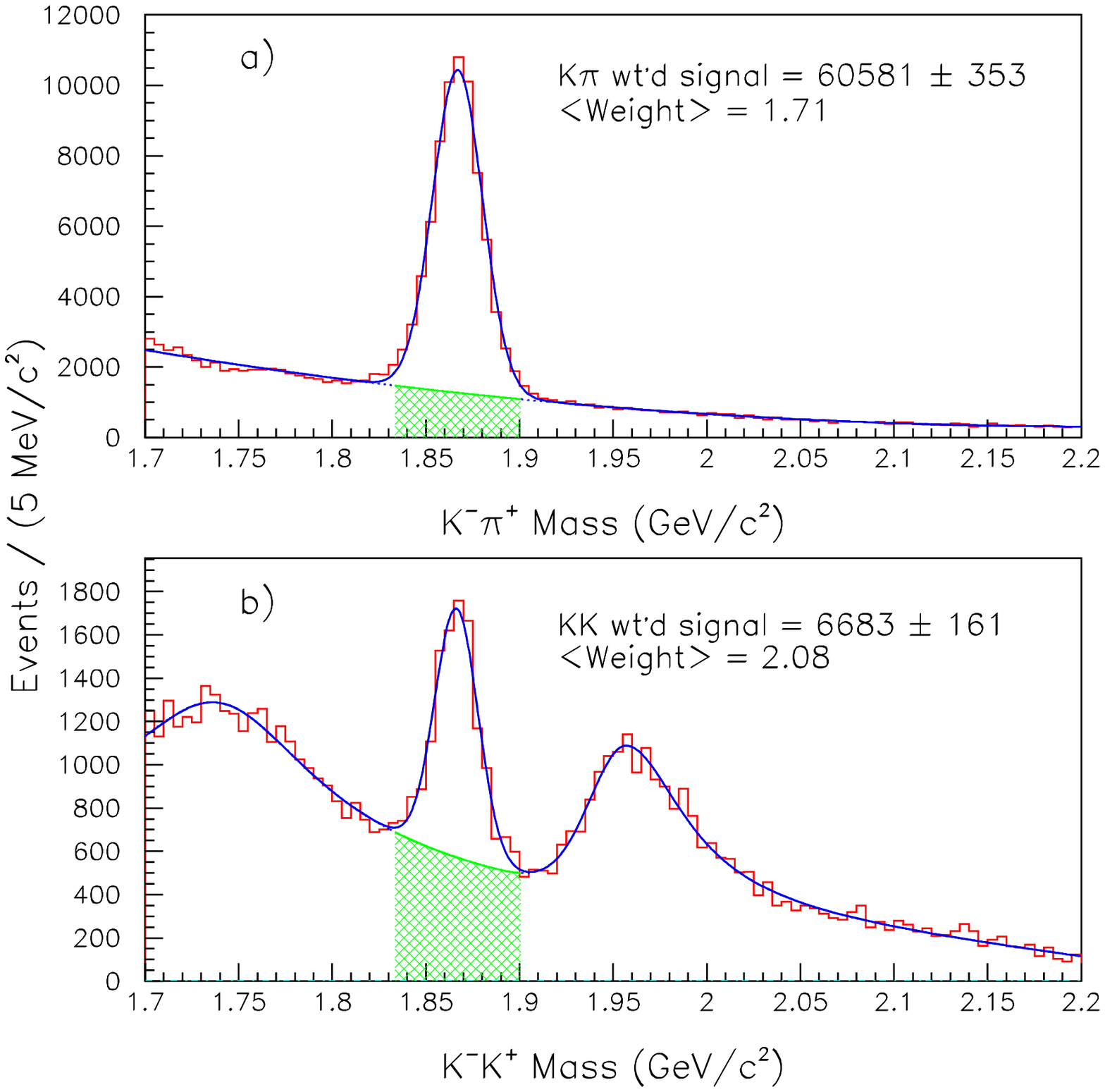,height=3.5in,width=2.4in}&
\epsfig{file=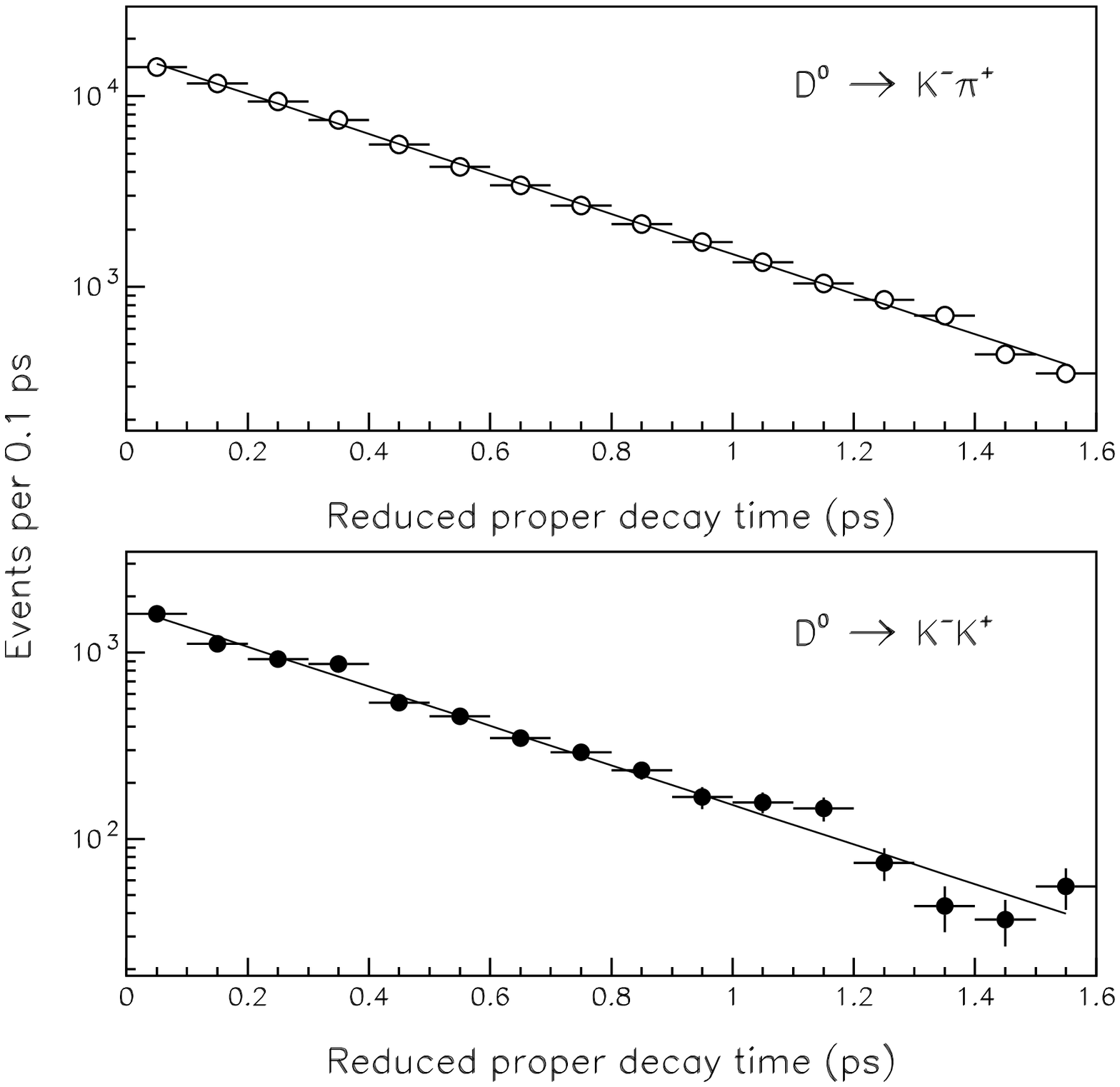,height=3.5in,width=2.4in}\\
\end{tabular}
\end{center}
\vspace{10pt}
\caption{Invariant mass for 
$D^0 \rightarrow K^- \pi^+$ (a) and $D^0 \rightarrow K^- K^+$ (b) decays 
(left). Exponential fits for number of decays as a function of 
reduced proper decay time (right), (E791 results).}
\label{tauD0}
\end{figure}

E791 has published searches for a lifetime difference between the $CP-even$ 
and $CP-odd$ eigenstates of the $D^0$ \cite{taud0-791}. 
To do so they compared  the lifetimes of the  decays 
$D^0 \rightarrow K^- K^+$ ($CP = +1$)  and $D^0 \rightarrow K^- \pi^+$, 
(CP mixed) \cite{cpvs}, shown in Fig. \ref{tauD0}.\\

\noindent
Defining 
\begin{equation}
\frac{\Gamma (K^- K^+) - \Gamma (K^- \pi^+)}{\Gamma (K^- \pi^+)} = y_{CP}
\end{equation}

\noindent
The time integrated ratio of mixed to non mixed decay rates in charm meson 
is given by

\begin{equation}
R_{mix}= \frac{\Gamma ( D^0 \rightarrow \overline {D^0} \rightarrow \bar {f})}
                {D^0 \rightarrow f} =  \frac{x^2 + y^2}{2} 
\end{equation}

\noindent
where
\begin{eqnarray}
 x = \frac{\Delta m}{\overline \Gamma}, \:\:\:
 y = \frac{\Delta \Gamma}{2 \overline\Gamma} \nonumber
\end{eqnarray}
 with 
\begin{eqnarray}
\Delta m = m_1-m_2, \:\:\:
\Delta \Gamma = \Gamma_1 - \Gamma_2,\:\:\:
\overline{\Gamma} = ({\Gamma_1 + \Gamma_2})/2 \nonumber
\end{eqnarray}

\noindent
where $\Gamma_1$ is for CP even states and $\Gamma_2$ for CP-odd states.\\
$\Gamma_1$ applies to $D^0 \rightarrow K^- K^+$ and
$\Gamma$ applies to $D^0 \rightarrow K^- \pi^+$ if CP is conserved.\\
Mixing can appear if there is either a difference in the masses of the CP 
eigenstates $\Delta m$ or if there is a difference in the decay rates 
$\Delta \Gamma$.\\

\noindent
E791 observed no difference in lifetimes and quoted
$\tau (K \pi)$ = $0.413 \pm 0.003 \pm 0.004$ ps 
 $\tau (KK)$ = $0.410 \pm 0.011 \pm 0.006$ ps, 
 $y_{CP}$ = $0.008 \pm 0.029 \pm 0.010$
 or $-0.04 < y_{CP} < 0.06$   ( $90 \% $  CL)
 or $\Delta \Gamma $= $2 ( \Gamma_{KK} - \Gamma_{K \pi})$ 
= $(0.04 \pm 0.14 \pm 0.05) \: ps^{-1}$

\section*{ACKNOWLEDGEMENTS}

One of us, J. C. Anjos would like to thank the conference organizers for 
their  invitation to attend the Symposium.
E. Cuautle  would like to thank the Centro Brasileiro de Pesquisas 
F\'{\i}sicas 
(CBPF) for its kind hospitality during his postdoctoral stay.
The authors would like to thank  CLAF/CNPq (Brazil) and CONACyT (M\'exico) for
financial support of this work.


\end{document}